%% file: main.tex
\documentclass[lettersize,journal]{IEEEtran}

\ifCLASSOPTIONcompsoc
  \usepackage[nocompress]{cite}
\else
  \usepackage{cite}
\fi

\usepackage{graphicx}
\usepackage{amsfonts} 
\usepackage{array}
\usepackage{lipsum}
\usepackage{subcaption}
\usepackage{longtable}
\usepackage{xspace}
\usepackage{algorithm}
\usepackage[noend]{algpseudocode}
\usepackage{amsmath}

\usepackage{subcaption}
\usepackage{relsize}
\usepackage{xcolor}
\usepackage{mathtools}
\usepackage{multirow}
\usepackage{marginnote}
\usepackage{pifont}
\usepackage{url}

\newcolumntype{L}[1]{>{\raggedright\let\newline\\\arraybackslash\hspace{0pt}}m{#1}}
\newcolumntype{C}[1]{>{\centering\let\newline\\\arraybackslash\hspace{0pt}}m{#1}}
\newcolumntype{R}[1]{>{\raggedleft\let\newline\\\arraybackslash\hspace{0pt}}m{#1}}

\newcommand{\VDA}[1]{$\mathbf{A_{12}}$}

\newcommand{\APPR}{DESIGNATE\xspace}
\newcommand{\APPRMONO}{DESIGNATE$_{single}$\xspace}
\newcommand{\APPRPIXEL}{DESIGNATE$_{pix}$\xspace}
\newcommand{\APPRNOGAN}{DESIGNATE$_{NoGAN}$\xspace}
\newcommand{\DEEPJANUS}{DeepJanus\xspace}
\newcommand{\DEEPJANUSGAN}{DeepJanus$_{GAN}$\xspace}

\definecolor{mygreen}{rgb}{0.0, 0.2, 0.13}

\newcommand{\MAJORBEGIN}{\color{black}}
\newcommand{\MAJOREND}{\color{black}}

\newcommand{\MAJOR}[2]{#2}

\begin{document}

\title{Search-based DNN Testing and Retraining with GAN-enhanced Simulations}

\author{Mohammed~Oualid~Attaoui, Fabrizio~Pastore, Lionel~C.~Briand
\IEEEcompsocitemizethanks{\IEEEcompsocthanksitem M. O. Attaoui and F. Pastore are with the SnT Centre, University of Luxembourg.\protect\\
E-mail: mohammed.attaoui@uni.lu, fabrizio.pastore@uni.lu
\IEEEcompsocthanksitem L. Briand is with the Lero Research Ireland Centre and University of Limerick, Limerick, Ireland, and the School of EECS, University of Ottawa, Ottawa, Canada.\protect\\
E-mail: lbriand@uottawa.ca}
}

\IEEEtitleabstractindextext{%
\begin{abstract}
In safety-critical systems (e.g., autonomous vehicles and robots), Deep Neural Networks (DNNs) are becoming a key component for computer vision tasks, particularly semantic segmentation. Further, since DNN behavior cannot be assessed through code inspection and analysis, test automation has become an essential activity to gain confidence in the reliability of DNNs. Unfortunately, state-of-the-art automated testing solutions largely rely on simulators, whose fidelity is always imperfect, thus affecting the validity of test results. To address such limitations, we propose to combine meta-heuristic search, used to explore the input space using simulators, with Generative Adversarial Networks (GANs), to transform the data generated by simulators into realistic input images. Such images can be used both to assess the DNN accuracy and to retrain the DNN more effectively. We applied our approach to a state-of-the-art DNN performing semantic segmentation, in two different case studies, and demonstrated that it outperforms a state-of-the-art GAN-based testing solution and several other baselines. Specifically, it leads to the largest number of diverse images leading to the worst DNN accuracy. Further, the images generated with our approach, lead to the highest improvement in DNN accuracy when used for retraining. In conclusion, we suggest to always integrate a trained GAN to transform test inputs when performing search-driven, simulator-based testing.
\end{abstract}

\begin{IEEEkeywords}
GAN-based testing, Simulator-based testing, DNN-based systems testing
\end{IEEEkeywords}}

\maketitle

\IEEEdisplaynontitleabstractindextext

\IEEEpeerreviewmaketitle

\input{introduction.tex}

\input{background.tex}
\input{approach.tex}
\input{evaluation.tex}
\input{related.tex}
\input{conclusion.tex}

\section*{Acknowledgments}
The experiments presented in this paper were carried out using the HPC facilities of the University of Luxembourg (see \url{http://hpc.uni.lu}). This work has been partially supported by the ESA contract RFP/3-17931/22/NL/GLC/my,
TIA (Test, Improve, Assure). Lionel Briand was partly supported by the Canada Research Chair and Discovery Grant programs of the Natural Sciences and Engineering Research Council of Canada (NSERC) and the Science Foundation Ireland under Grant 13/RC/2094-2. 
\bibliographystyle{plain}
\bibliography{bib}

\end{document}

%% file: introduction.tex
\section{Introduction}
\label{sec:introduction}
Deep Neural Networks (DNNs) are an effective solution to automate tasks that require vision capabilities. For this reason, several companies developing cyber-physical systems (CPS), 
are making large R\&D investments to rely on DNNs to automate tasks that are currently manual. Examples in the automotive sector include driving assistance systems (e.g., obstacles detection, traffic sign detection, autonomous driving) and interior car monitoring (e.g., drowsiness detection or unsupervised child detection). In other sectors, key developments are observed in robotics and space, for example to develop autonomous vehicles capable of exploring caves and planets' surfaces. 

When a DNN that processes camera images is used to drive a CPS, we must gain sufficient confidence that it can adequately respond to all foreseeable inputs. 
To achieve this objective, research has focused on combining meta-heuristic search with simulators~\cite{Haq:2021,fitash:offline:emse,Gladisch2019,GambiASE19,riccio2020model,zohdinasab2021deephyperion}.  The rationale is that, under the assumption that simulators can accurately create images corresponding to a realistic scene, meta-heuristic search enables the cost-effective search of inputs leading to failures.  What enables test automation is the fact that simulators, in addition to generate images according to the parameters selected by the search algorithm, can be used to automatically assess the quality of the DNN output for each generated input. For example, the performance of steering angle DNNs can be assessed by relying on the distance between the ego vehicle and the lane separator line, both provided by the simulator; indeed, the closer the vehicle gets to the separator line, the more likely it will cross the lane. Although such search-based approaches effectively generate images leading to DNN failures, the validity of such DNN assessment highly depends on the simulator fidelity. For example, recent work has demonstrated that the fidelity gap of simulators leads to different reliability results in simulation-based and real-world testing~\cite{fitash:offline:emse,stocco2022mind}. 

The limited fidelity of a simulator may thus hinder the detection of DNN shortcomings.
For example, the road landscapes generated by the AirSim simulator~\cite{shah2018airsim}, which resemble residential areas in the USA, are very different from
the landscape of European cities and towns. 
 Based on our interactions with industry partners in the space and automotive sectors~\cite{fahmysupporting,Hazem:SEDE,Mohammed:PipelinesAssessment}, we  noticed that high-fidelity simulators are lacking for many environments in which DNNs can be applied, including for example car interiors (e.g., for in-cabin monitoring DNNs)  and space landscapes (e.g., to drive Mars and Lunar rovers).

To accurately test DNNs, ensuring that we explore the input space relying on realistic input images, we propose 
\emph{\underline{D}NN t\underline{es}ting w\underline{i}th  \underline{G}AN-e\underline{n}hanced simul\underline{at}ions and s\underline{e}arch (\APPR)}.
We rely on meta-heuristic search to effectively drive a simulator towards the generation of diverse inputs leading to failures but, instead of testing the DNN with simulator images, 
we leverage Generative Adversarial Networks (GANs) to transform the simulator output into a more realistic image that resembles real-world data distributions.
GANs are a promising candidate to address the simulator fidelity gap because they have shown to be effective in image-to-image translation, that is mapping images from one domain (e.g., sketches) to images in another (e.g., photographs)~\cite{MUNIT,wang2018pix2pixHD}. For testing, GANs  have been mainly adopted to introduce realistic perturbations in input images~\cite{zhang2018deeproad,li2021testing,Wei:AdversarialStyle} and to estimate sensory data~\cite{SurfelGAN,stocco2022mind}.  \APPR is the first technique integrating GANs into search-based testing with simulators.

We assume that, because the realistic images are generated from the simulated ones, both the simulator and realistic images share the same ground truth (i.e., the DNN should generate similar outputs from them), thus enabling test automation by assessing the DNN output based on the information provided  by the simulator. Further, realistic images leading to DNN failures  can be used to retrain a DNN to improve its reliability in the field, something which may be more difficult to achieve with simulator images because of 
on their fidelity. However, to address this challenge, \APPR can test a road scene segmentation DNN by relying on AirSim to generate landscapes of city roads with traffic and buildings, and then leverage a GAN to transform those landscapes into realistic images with European cities.

Inspired by work on DNN testing~\cite{fitash:offline:emse,riccio2020model}, we rely on genetic algorithms to cost-effectively identify failure-inducing images with GANs; our objective is to identify simulator parameters leading to simulator outputs (e.g., an image of a simulated city road) that, after being transformed into realistic images using a GAN, lead to DNN failures. Since we ideally aim at identifying all the situations in which the DNN fails, to both identify failures and maximize input diversity, we combine two fitness functions.  
The first one measures the accuracy of the DNN output whereas the second one measures the similarity across generated realistic images.
To test a DNN with diverse scenes being depicted in the input images, we are interested in generating images that differ in terms of the items being displayed and not just a few, scattered, unmatched pixels. For this reason, we introduce a feature-based distance metric that leverages transfer learning~\cite{Mohammed:PipelinesAssessment}.

We conducted an empirical investigation with a state-of-the-art, third-party DNN for road segmentation. Our results show that \APPR outperforms TACTIC~\cite{li2021testing} and \DEEPJANUS~\cite{riccio2020model}, two state-of-the-art DNN testing approaches, and several baselines including random search, single objective search, and multi-objective search without GANs. \APPR generates inputs leading to the worst DNN reliability and  maximizes input diversity. Further, we demonstrate that the inputs generated by \APPR better support DNN retraining, compared to competing approaches, with an increase in DNN accuracy of eight percentage points. \MAJOR{R3.6}{In addition to our road segmentation case study, in the context of the Mars rover self-driving system, we also demonstrate the applicability and benefits of our approach to object detection on Mars' surface, using the AI4MARS dataset \cite{swan2021ai4mars} and a custom-built simulator leveraging Unreal Engine~\cite{unrealengine} and Airsim.} Based on our results across both case studies, we conclude that GANs should be combined with simulators when testing and improving vision-based DNNs as it leads to more effective and diverse testing, as well as better retraining.


Our paper proceeds as follows. Section~\ref{sec:background} introduces background techniques. Section~\ref{sec:approach} describes \APPR. Section~\ref{sec:evaluation} reports on our empirical study. Section~\ref{sec:related} discusses related work. Section~\ref{sec:conclusion} concludes the paper.

%

%% file: background.tex
\section{Background}
\label{sec:background}

\subsection{Terminology}

In this paper, we rely on the following definitions:

\emph{A \textbf{scene} describes a snapshot of the environment including the scenery and dynamic elements, as well as all actors’ and observers’ self-representations, and the relationships among those entities}~\cite{Ulbrich:Terminology}. In our context, where the DNN under test processes pictures taken by a camera, a scene is a single picture taken by a camera or generated by a simulator. 

\begin{figure}
    \centering
    \includegraphics[width=\linewidth]{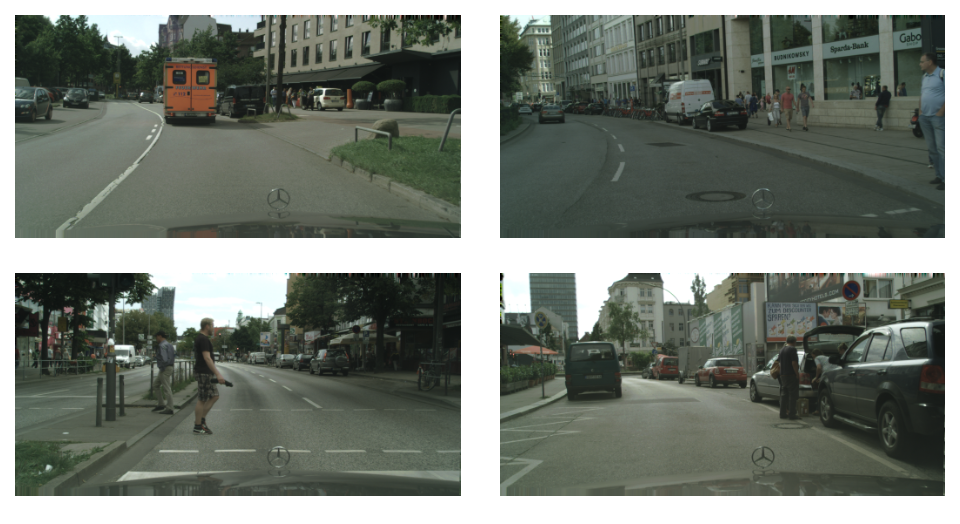}
    \caption{Example of images from the cityscapes dataset showing the same situation (a turning road, buildings on the side along with parked cars).}
    \label{fig:example_road}
\end{figure}

\emph{A \textbf{situation} is the entirety of circumstances, which are to be considered for the selection of an appropriate behavior pattern at a particular point of time~\cite{Ulbrich:Terminology}.} 
For vision DNNs in autonomous driving (e.g., semantic segmentation, steering angle prediction), since the DNN prediction depends on the objects depicted in an image, a situation captures commonalities among similar images leading to the same DNN prediction. For example, pictures with a turning road, buildings on the side along with parked cars, may belong to the same situation (see Figure~\ref{fig:example_road}).

\emph{A \textbf{scenario} describes the temporal development between several scenes in a sequence of scenes. Every scenario starts with an initial scene. Actions and events characterize the temporal development in a scenario. Other than a scene, a scenario spans a certain amount of time~\cite{Ulbrich:Terminology}.} In our context, a scenario is a sequence of images generated by the simulator.

\subsection{Semantic Segmentation}
\label{sec:segmentation}
Semantic segmentation is a computer vision task that involves assigning a class label to each pixel in an image. Unlike other forms of image recognition or classification, semantic segmentation provides a fine-grained understanding of an image's content by labeling each pixel with a corresponding class, such as ``road,'' ``car,'' ``person,'' and ``building,''.

Semantic segmentation is often performed with convolutional neural networks (CNNs) that learn to capture spatial relationships, object boundaries, and contextual information within an image, enabling them to differentiate between different objects and delineate their precise outlines.
In our study, we consider the state-of-the-art DeepLabV3+ architecture \cite{chen2018encoder}. DeepLabV3+ is renowned for its accuracy and efficiency in pixel-level semantic segmentation tasks \cite{liu2024image, wang2023dpnet}. 


\subsection{Simulated Environments}
\label{sec:airsim}

Simulators replicate real-world scenarios, allowing DNNs to be tested in a controlled and reproducible environment. 
In our work, we rely on two simulators: 
\subsubsection{AirSim}
AirSim is an open-source simulator for autonomous vehicles and drones developed by Microsoft~\cite{shah2018airsim}. It provides a realistic environment for testing and training AI algorithms using a variety of sensors, including cameras, lidars, and GPS. 
AirSim can be controlled through a dedicated API, which enables its use for DNN testing. In our context, we rely on the features for  controlling the positioning of the ego vehicle in the simulated environment and generating a picture of the landscape in front of the ego vehicle.
Further, for each generated image, AirSim provides a semantic segmentation image where each pixel's color directly corresponds to a specific category or class, including roads, buildings, vehicles, pedestrians, or sky.  We use these images as \emph{ground truth} to assess the reliability of DNNs in road scene segmentation.
\subsubsection{MarsSim}\MAJOR{R3.6}{
Focusing on extraterrestrial object detection, specifically on the Martian surface, we built a simulator that generate images resembling the ones in the AI4MARS dataset by combining Unreal Engine with the the same plugins that Airsim uses. This simulator creates realistic images of Mars terrain, which, when enhanced with a GAN, enables the generation of diverse training data for testing DNNs in Mars exploration contexts. Including the Airsim API enables us to control the simulator by changing the position of the camera. 
To create an authentic Martian environment, we modeled objects that match those found in the AI4MARS dataset: rocks, bedrock, sand, and Martian soil, each designed with textures and colors reflective of actual Mars imagery. We also incorporated variations in object size, shape, and placement to reflect the irregular distribution of terrain features typically observed on Mars. Additionally, we introduced terrain undulations and slopes to simulate the topographical challenges of Mars, such as craters, cliffs, and valleys, which pose unique navigational challenges for autonomous rovers.}

\subsection{GAN-based Image-to-image Translation }


Image-to-image translation maps images from one domain (e.g., sketches, edges, or semantic labels) to images in another domain (e.g., photographs, paintings, or architectural layouts). This transformation process involves learning the complex and non-linear mappings between the two domains, allowing for a seamless transition while preserving essential attributes. 

Generative Adversarial Networks (GANs) are a key technology for image-to-image translation.
GANs consist of two interconnected DNNs - the generator and the discriminator. The generator is responsible for producing synthetic images, while the discriminator aims to differentiate between real images from the target domain and the generated images. Through a competitive process, the generator progressively improves its ability to produce more realistic images, and the discriminator becomes more adept at distinguishing real and generated images.

\begin{figure}[h]
    \centering
    \includegraphics[width=0.5\textwidth]{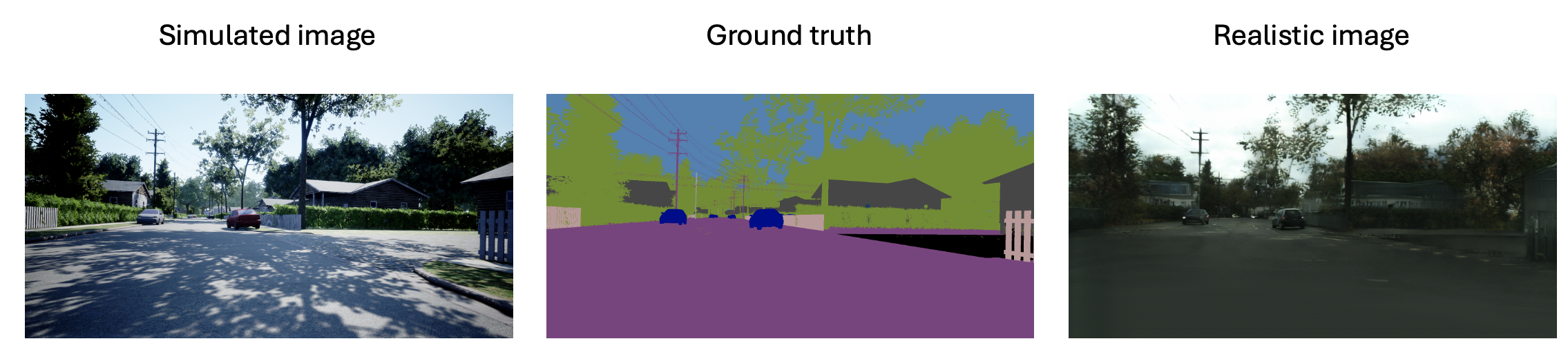}
    \caption{Examples of simulator image, ground truth (i.e., segmentation map provided by the simulator), and realistic image generated by Pix2PixHD from the ground truth of AirSim.}
    \label{fig:example}
\end{figure}

Example GANs for image-to-image translation are Pix2Pix~\cite{isola2017image}, with its high definition extension Pix2PixHD~\cite{wang2018pix2pixHD}, which are trained with pairs of images from the two domains, and CycleGAN~\cite{Zhu:2017}, which does not require paired images. Other GANs are UNIT~\cite{Liu:UNIT} and MUNIT~\cite{MUNIT}. UNIT includes two domain image encoders, two domain image generators, and two domain adversarial discriminators. MUNIT decomposes an image representation into domain-independent content code and domain-specific style code. The content code captures the underlying structure and semantics of the image that remain consistent across different domains. For example, in road scene images, the content code includes the road layout, lane positions, vehicles, traffic signs, pedestrians, and buildings. On the other hand, the style code captures the stylistic attributes unique to a particular domain. In the context of automotive images, style codes might include night-vision images, different weather conditions, and synthetic images from simulators. MUNIT translates images to other domains by recombining content code with style code sampled from the target domain. Related work leveraged UNIT and MUNIT to change weather conditions in real-world images~\cite{AdversarialStyle,zhang2018deeproad}, which is a different objective than ours. CycleGAN has instead been used to automatically generate sensor data~\cite{stocco2022mind}. After a preliminary assessment of both CycleGAN and Pix2PixHD to generate realistic images from segmentation maps produced by the AirSim simulator, we selected Pix2PixHD for our work since it led to better results (i.e., higher-fidelity images).
Figure \ref{fig:example} illustrates an example of image-to-image translation of a segmentation map produced by AirSim into a realistic image, using the Pix2PixHD model~\cite{isola2017image}

\begin{figure}[h]
    \includegraphics[width=8.8cm]{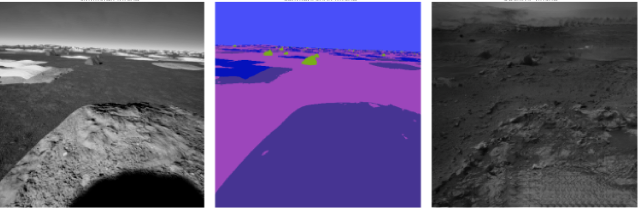}
    \caption{Examples of a Mars simulator's simulated image, its ground truth, and the realistic image generated from it by Pix2pixHD.}
    \label{fig:example_mars}
\end{figure}

%% file: approach.tex
\section{The \APPR Approach}
\label{sec:approach}
\APPR supports the testing and improvement of vision DNNs by generating realistic but synthetic inputs belonging to distinct situations. Such realistic inputs are generated by combining meta-heuristic search, simulators, and GANs. Simulators enable the inexpensive generation of input images and corresponding ground truth information (e.g., correct segmentation mask),
while meta-heuristic search enables the exploration of the input space to exercise distinct situations increasingly likely to lead to mispredictions. Finally, GANs enable the generation of high-fidelity images from the simulator data, thus maximizing the likelihood that testing results are representative of what can be observed with real-world inputs, thus ensuring that the mispredictions observed during testing are due to DNN limitations that can be experienced in the real-world and not the fidelity gap of simulators. Further, realistic images can be used to retrain the DNN under test to improve its reliability. Last, the images generated by \APPR belong to distinct situations,
thus enabling the identification of diverse DNN limitations and their improvement. 

\begin{figure*}[t]
    \centering
    \includegraphics[width=\textwidth]{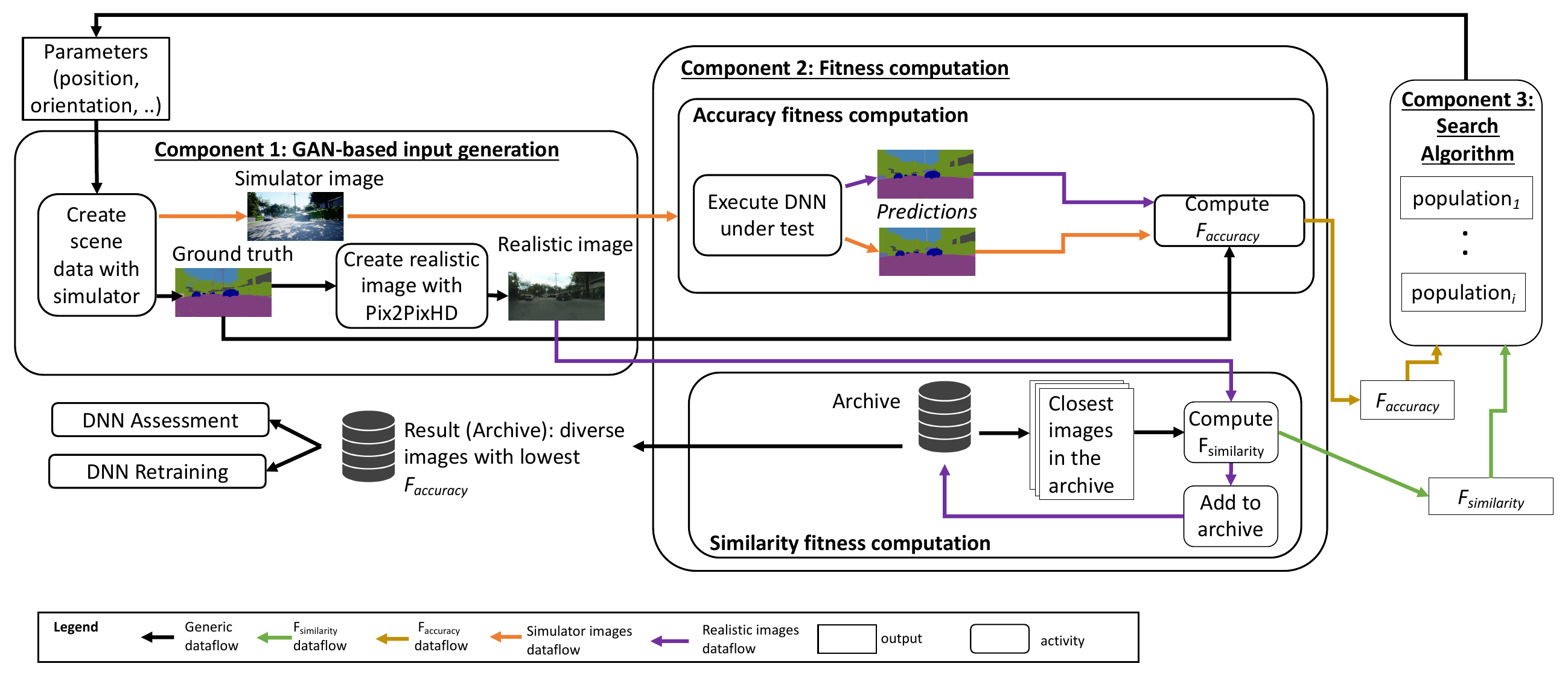}
    \caption{Overview of \APPR.}
    \label{fig:approach}
\end{figure*}

Figure \ref{fig:approach} provides an overview of \APPR, which consists of three components: 
\begin{enumerate}
\item \emph{GAN-based input generation component.} It receives as input the configuration parameter values to be used to generate, with a simulator, \emph{scene data} (i.e., an image and the corresponding segmentation map). The scene data is then used to generate a \emph{realistic image} with a GAN trained on a specific reference domain. For our automotive case study, we used the Pix2PixHD GAN trained on the CityScape dataset to generate realistic road scenes from the segmentation map produced by the simulator. \MAJOR{R3.6}{For our Mars case study, we utilized the AI4MARS dataset to train the GAN, and the simulator was built using Unreal Engine and Airsim to generate Martian surface images.}

\item \emph{Fitness computation component.} Since we aim to identify situations that are both unsafe (i.e., leading to DNN mispredictions) and diverse, we apply two distinct fitness functions: $F_{accuracy}$, which measures how accurate is the DNN prediction on the generated realistic image (e.g., the accuracy of the semantic segmentation), and $F_{similarity}$, which measures how similar a generated image is from the already generated ones.  These functions are used by a minimization algorithm to address our objectives, as described next.

\item \emph{Search Algorithm.} We rely a multi-objective search algorithm to find a population of diverse situations where the DNN fares poorly, thereby suggesting these scenarios are not well captured by the real-world DNN training set. In our search algorithm, an individual is captured by the set of parameters used to deterministically generate, using a simulator, scene data, and to derive a realistic image from such data. For optimization purposes, for each individual, we store the simulator parameter values, the scene data, and the realistic image.

\end{enumerate}

The output of \APPR is a set of diverse and realistic images leading to poor DNN reliability. They should be inspected for safety-analysis purposes; specifically, they enable understanding the situations in which the DNN mispredicts, determine their likelihood based on domain knowledge, and identify countermeasures. Also, they can be used to retrain the DNN and improve its performance. The following sections provide details about our components.

\subsection{GAN-based input generation}
\label{sec:approach:inputGeneration}
We generate scene data (i.e., images with simulated scenes and corresponding segmentation masks) by relying on the AirSim simulator (Section \ref{sec:airsim}); however, our approach can be applied with simulators having similar capabilities (e.g., Carla~\cite{carla} and BeamNG~\cite{beamng}). However, one must ensure that the generated scenes are realistic. In our case, to assess DNNs that control a vehicle (e.g., segmentation of the images in front of the vehicle, or steering angle prediction), it is necessary to generate scenes that mimic what can be observed by a car on the road. Consequently, we control the generation of scene data through two parameters: orientation and position of the car.
 To test other DNNs (e.g., visual navigation for drones), a different set of parameters would be considered (e.g., altitude might be included).


We then generate realistic images from the ground truth information provided by the simulator. In our experiments, such ground truth is the segmentation mask, but could be different information in other contexts.
To this end, we utilize a Pix2PixHD DNN~\cite{wang2018pix2pixHD} that we train using the same dataset used for the DNN under test; such dataset consists of pairs $\langle \text{image}, \text{expected output} \rangle$. For example, in our experiments, we considered a DNN trained on a dataset of pairs $\langle \text{European road image}, \text{segmentation mask} \rangle$, which we use to train Pix2PixHD to generate images of roads belonging to European cities from segmentation masks.

During testing, \APPR provides the ground truth of the simulated images to the trained Pix2PixHD to generate a realistic images; in our experiments, \APPR provides to Pix2PixHD the segmentation mask produced by AirSim in order to generate realistic images resembling real-world images from Cityscapes.





\subsection{Fitness Functions}

\subsubsection{Accuracy fitness}
\label{approach:accuracyFitness}

The fitness function $F_{accuracy}(i)$, which computes the accuracy of the DNN on an individual, depends on the DNN under test. In general, we suggest relying on the metric used to measure the reliability of the DNN. For example, in our empirical assessment, we test a DNN for image segmentation that is assessed, for safety reasons, based on its capability to detect cars accurately. Specifically, it is assessed with the \emph{Intersection over Union (IoU)} metric, (i.e., Jaccard Index~\cite{everingham2010pascal}). It measures the overlap between the predicted segmentation mask and the ground truth mask for a class (e.g., car, road), as the ratio of the intersection area to the union area of the two masks. 
 Therefore, we compute $F_{accuracy}(i)$ by relying on the IoU formula for car objects:
 
\begin{equation}
IoU_{car}(i) = \frac{{TP_{car}(i)}}{{TP_{car}(i) + FP_{car}(i) + FN_{car}(i)}}
\end{equation}

where $TP_{car}(i)$ are the true positive car pixels in image $i$ (pixels that belong to cars in both ground truth and prediction), $FP_{car}(i)$ are the false positive pixels for cars (pixels erroneously reported as belonging to a car in the predicted segmentation mask), and $FN_{car}(i)$ are the false negative pixels (pixels erroneously not reported as belonging to a car in the predicted segmentation mask).

\MAJOR{R3.6}{For the Martian case study, we consider all the classes when computing the accuracy. Therefore, we use the mean IoU ($mIoU$) as $F_{accuracy}(i)$. The $mIoU$ for one image is obtained by averaging the IoU scores across all classes and is computed as follows:}

\begin{equation}
mIoU = \frac{1}{C} \sum_{c=1}^{C} \frac{TP_c}{TP_c + FP_c + FN_c}
\end{equation}

Unfortunately, when test images are automatically generated using simulators and GANs, the DNN may badly perform because the generated images are not relevant for testing. Specifically, they may be out-of-domain (because of the simulator's control limitations) or may not match the ground truth (because of the GAN's limitations). 
Consequently, we defined a fitness function that distinguishes between non-relevant and relevant individuals, as follows:

\begin{equation}
F_{accuracy} (i) =  
    \begin{cases}
      2 & \text{if } i \text{ is not relevant}\\
      \mathit{performance}(i) & \text{otherwise}
    \end{cases}   
\end{equation}

We assume the performance metric (i.e., $performance(i)$, based on $IoU{car}(i)$) to be normalized in the range 0-1, with a perfect output leading to $1$. Therefore, since in our search we aim to minimize the fitness function, we return a fitness of 2 for irrelevant individuals, so that they are discarded by the search process. Below, we describe how we determine irrelevant individuals.

How we determine \emph{out-of-domain} inputs depends on the specific case study.
When relying on AirSim, for example, we observe out-of-domain inputs when the ego vehicle is positioned next to another vehicle and perpendicular to it (e.g., facing the other vehicle's door), which is not interesting for our testing purposes since it is very easy to make a segmentation DNN fail in such case.
Another case is that of images with no cars, which should be excluded because we are only interested in testing the DNN with images exhibiting traffic. Consequently, we consider an image to be relevant if the proportion of pixels  belonging to cars (i.e., $\mathit{carPixelsProportion}$) in the input image is within a given range:
$$0<\mathit{carPixelsProportion}<0.4$$ 
\MAJOR{R3.2}{The variable \emph{carPixelsProportion} is  computed on the segmentation ground truth. We set the upper bound to $0.4$ to ensure that the car does not cover most of the image, which would result in unrealistic scenarios for evaluating DNN performance. Indeed, images where \emph{carPixelsProportion} exceeds 0.4 often involve close-ups or unusual viewpoints, such as a vehicle occupying most of the camera's field of view---lidar-based sensors typically prevent autonomous cars from being in such situations.} 
We set the lower bound above zero to ensure that the image contains at least one car. 
\MAJOR{R3.6}{Similarly, for the Martian environment, we need to ensure that the sky does not cover most of the image and, therefore, we 
consider an image relevant if the proportion of pixels  belonging to the sky (i.e., $\mathit{skyPixelsProportion}$) is within the range:
$$0<\mathit{skyPixelsProportion}<0.7$$ 
One may consider that $70\%$ is a relatively large proportion of the image but it is a realistic situation based on our discussions with ESA representatives.}


Also, to prevent the positioning of the car outside of the road (e.g., on a building), we ensure that the position of the car is within the road lanes, whose coordinates we mapped manually, by assigning a high fitness otherwise.

Detecting \emph{GAN-generated realistic inputs not matching the ground truth} is hard to automate because only humans can determine if the image is a correct representation of the ground truth (e.g., cars really appear where car pixels are reported on the ground truth).
However, to automate such verification, we propose relying on heuristics based on the difference in DNN reliability between images generated by the simulator and the corresponding GAN-generated realistic images. 
Our heuristic relies on the empirical observation that the DNN under test performs better on a realistic image than on the corresponding simulator image, which is expected with DNNs trained on real-world images. 
To apply our heuristic, before testing, we consider a large set  (1000) of simulator images  generated with randomly selected parameter values, and generate corresponding realistic images.  Each simulator and realistic image is processed by the DNN under test to generate a prediction, and the performance metric is computed for each image in each image pair, which enables computing the performance difference as: 
\begin{flalign}
    & \mathit{delta}_{performance}(i) =  \\ \nonumber 
    & \mathit{performance}_{simulated}(i) - \mathit{performance}_{realistic}(i)
\end{flalign}

where $i$ indicates the i-th pair of simulator and corresponding realistic image.

After sorting the generated image pairs based on their $\text{delta}_{performance}(i)$, it is possible, through visual inspection, to determine a threshold ($T_{relevance}$) above which the realistic images do not correctly capture the ground truth (e.g., a car is missing where expected based on the ground truth). We selected $T_{relevance}$ as the third percentile of the $\mathit{delta}_{performance}$ distribution.
During testing, we thus consider a realistic image to be relevant if ${delta}_{performance}(i) \le T_{relevance}$. Regarding the segmentation DNN for urban environments considered in our empirical assessment:

    \begin{flalign}
        & \mathit{delta}_{performance}(i) = \mathit{delta}_{IoU}(i) = \\ & \nonumber\mathit{IoU}_{simulated}(i) - \mathit{IoU}_{realistic}(i)
    \end{flalign}

 \MAJOR{R3.6}{In contrast, in the case of the martian environment, we did not observe GAN-generated images that are less realistic than those of the simulator and therefore we do not rely on $T_{relevance}$. We believe this is mainly due to our Mars simulator having limited fidelity and is thus representative of cases where limited effort was dedicated to simulator development\footnote{Please note that state-of-the-art realistic simulators for space exist (e.g., PANGU~\cite{PANGU}), but they do not generate the ground truth labels required to train segmentation DNNs.}.}



In the segmentation DNN considered in our empirical assessment with the urban case study, $F_{accuracy}$ is therefore computed as follows:
\begin{equation}
\label{eq:f1}
\small
F_{accuracy} (i) =  
    \begin{cases}
      2 & \text{if } {0}<\mathit{carPixelsProportion}<{0.4}\\
      2 & \text{if } \text{the ego vehicle is off the road}\\
      2 & \text{if } {delta_{IoU}(i) > T_{relevance}}\\
      IoU_{car}(i) & \text{otherwise}
    \end{cases}   
\end{equation}

\MAJOR{R3.6}{For the martian case study, based on the above discussion, $F_{accuracy}$ is instead computed as follows:}
\MAJORBEGIN
\begin{equation}
\label{eq:f1}
\small
F_{accuracy} (i) =  
    \begin{cases}
      2 & \text{if } {0}<\mathit{skyPixelsProportion}<{0.7}\\
      mIoU(i) & \text{otherwise}
    \end{cases}   
\end{equation}
\MAJOREND


\subsubsection{Diversity fitness}

Our diversity fitness should enable determining if a newly generated image captures a situation that was not observed before (e.g., no traffic on the road VS a lot of traffic on the road), which can be achieved by computing the difference between the newly generated image and the previously generated images (e.g., stored in an archive). However, the above-mentioned objective cannot simply rely on a simple computation of the  the pixel-by pixel difference between two images; indeed, two images generated before and after slightly turning a car, may present pixels that are all shifted in one direction thus resulting into a high difference even if the two images capture the same situation. 

To address the problem above, we suggest capturing the semantic difference between two images by computing the difference between the feature vectors derived from the two images.
Based on our previous experience with clustering DNN mispredictions~\cite{Mohammed:PipelinesAssessment}, we rely on a ResNet50 model trained on the ImageNet dataset to automatically derive feature vectors. The deep architecture of ResNet50 (50 layers) allows it to capture richer and more abstract features at different levels of the network. This makes it more powerful for feature extraction tasks where high-level semantic features are important~\cite{he2016deep, simonyan2015very}. 
Image diversity can then be computed as the Euclidean distance between two feature vectors.
Euclidean distance inherently measures the geometric distance between points in a multi-dimensional space; when applied to feature vectors representing images, it captures the differences in features between images. Images with similar features will have a shorter distance, indicating less diversity, while images with dissimilar features will have a longer distance, suggesting higher diversity. 

In the presence of an archive with previously generated images, $F_{similarity}$, which captures how $i$ is similar to previously generated images, can thus be computed as a function of the Euclidean distance from the closest image in the archive. However, in a search algorithm, preserving images that slightly differ from previously generated ones may lead to being stuck in local optima; therefore, inspired by previous work \cite{riccio2020model}, we introduce a threshold for the diversity ($T_{diversity}$) below which an image is not considered different from a previously generated one and is thus assigned with high fitness:

\begin{equation}
\label{eq:f2}
\begin{aligned}
F_{similarity} (i) =  
    \begin{cases}
      2 ~~ \text{if } \mathit{distanceFromClosest}(i) < T_{diversity}\\
      \frac{1}{1 + \mathit{distanceFromClosest}(i))} ~~ \text{otherwise}
    \end{cases}   
\end{aligned}
\end{equation}

The value of $\mathit{distanceFromClosest}{(i)}$ is identified by computing the Euclidean distance of the image $i$ from each image in the archive and selecting the shortest one. 
Like in related work~\cite{riccio2020model}, we compute $T_{diversity}$ as the median of the pairwise distances in a set of 1000 randomly generated images. 
Finally, the term $\frac{1}{1 + \mathit{distanceFromClosest}_(i)}$ is used for normalization: it leads to a fitness close to zero for a very high distance, and to one for overlapping images.



\subsection{Search algorithm}
\label{sec:genetic}
We aim to address a multi-objective problem \cite{attaoui2020multi}, generating realistic images leading to minimal DNN accuracy and having minimal similarity among each other. We therefore rely on the NSGA-II algorithm, which demonstrated its effectiveness in multiple multi-objective software testing problems. However, inspired by related work aiming at maximizing the diversity among the generated individuals~\cite{riccio2020model}, 
we extend NSGA-II with an archive to store the best (non-dominated) individuals encountered during the search process. The archive plays a critical role in preventing the genetic algorithm from cycling, a phenomenon where the population moves from one area of the solution space to another and then back again without exploring other areas.
By retaining the best individuals in the archive, and assigning high fitness values to individuals that are similar to the already generated ones, the algorithm ensures that these areas are not revisited, facilitating a more efficient exploration of the solution space.


\begin{algorithm}[t]
\footnotesize
\caption{Updating the archive with image i}
\label{alg:update}
\begin{algorithmic}[1]
\Require $A$: archive with individuals selected so far, $i$: individual to be added, $T_{similarity}$
\Ensure  archive with individuals that are sparse and minimize $F_{accuracy}$
\If{$F_{accuracy}(i) == 2$} \Comment{ignore non-relevant}\label{algo:archive:ignore:unrealistic}
   \State \textbf{return}
\EndIf   
\If{$\mathit{distanceFromClosest}(i) > T_{diversity}$} 
   \State add i to $A$ \label{algo:archive:include:distant}
\ElsIf{$\big(\mathit{distanceFromClosest}(i) \leq T_{diversity}$ \textbf{and} $F_{accuracy}(closestTo(i)) > F_{accuracy}(i)\big)$}
  \State put $i$ in $A$, in place of i's closest individual already in $A$ \label{algo:archive:include:lowerAcc}
\EndIf
\end{algorithmic}
\end{algorithm}

Before delving into our search algorithm, we describe the procedure adopted to update the archive with an individual $i$, which is shown in Algorithm~\ref{alg:update}. 
Individuals that are not relevant for testing (i.e., $F_{accuracy}(i) == 2$) are not added to the archive (Line~\ref{algo:archive:ignore:unrealistic}). Instead, we add individuals that are relevant and whose distance from the closest individual in the archive is above the predefined threshold (Line~\ref{algo:archive:include:distant}).
Last, since individuals that are similar to others in the archive (i.e., $\mathit{distanceFromClosest}(i) < T_{diversity}$) may still be more effective for testing (i.e., lower $F_{accuracy}$), we put the individual $i$ in place of the individual that is closest to $i$ in the archive when the $F_{accuracy}$ of the former is lower than the accuracy of the latter (Line~\ref{algo:archive:include:lowerAcc}).


\begin{algorithm}[t]
\caption{Search algorithm used in \APPR}
\label{alg:search}
\begin{algorithmic}[1]
\footnotesize

\Require $g_{max}$: maximum number of generations, $N$: population size, $P_c$: crossover probability, $P_m$: mutation probability, $r$: number of populations for initial seed selection
\Ensure An archive of diverse individuals leading to minimal DNN accuracy 
\State Initialize an empty archive $A$ \label{algo:search:createArchive}
\State Initialize $r$ random populations $P^0_0$..$P^r_0$ with $N$ individuals in each population \label{algo:search:start}
\State Evaluate $F_{accuracy}$ and $F_{similarity}$ of each individual in each population $P^0_0$..$P^r_0$ based on Equations~\ref{eq:f1} and~\ref{eq:f2}
\State Select $P_0$ as the population with the individual having the lowest $F_{accuracy}$. \label{algo:search:select:start}
\State Update the archive $A$ based on $F_{accuracy}$ and $F_{similarity}$ of each individual in $P_0$ \label{algo:search:firstUpdate}
\State Rank the individuals in  using non-dominated sorting \label{algo:search:sort}
\State Calculate crowding distance for each individual \label{algo:search:crowd}
\State $g \gets 0$
\While{$g < g_{max}$}
    \State Select parents from the population $P_g$ using binary tournament selection based on rank and crowding distance \label{algo:search:parents}
    \State Generate offspring $Q_g$ using polynomial mutation and simulated binary crossover \label{algo:search:offspring}
    \State Combine $P_g$ and $Q_g$ into $R_g$
    \State Compute $F_{accuracy}$ for each individual in $Q_g$ \label{algo:search:compute:metrics}
    \State Compute $F_{similarity}$ for each $i$ in $R_g$\label{algo:search:loop:diversity}
    \For{$i$ in $R_g$}
    \State Update the archive $A$ with $i$ using on Algorithm~\ref{alg:update}  \label{algo:search:loop:update}
    \EndFor
    \State Update $F_{similarity}$ of each individual in $R_g$
    \label{algo:search:loop:update_diversity}
    \State Rank $R_g$ using non-dominated sorting \label{algo:search:loop:rank}
    \State Calculate crowding distance for each individual in $R_g$ \label{algo:search:loop:crowd}
    \State Select the top $N$ individuals from $R_g$ to form new population $P_{g+1}$ based on rank and crowding distance \label{algo:search:loop:select}
    \State $g \gets g + 1$
\EndWhile
\State \textbf{return} $A$

\end{algorithmic}
\end{algorithm}

Algorithm \ref{alg:search} presents the pseudocode of the NSGA-II algorithm (with archive) employed in \APPR. After creating an empty archive (Line~\ref{algo:search:createArchive}),
since the initial selection of seeds may affect the search results and inspired by our previous work~\cite{SAFE}, we generate multiple ($5$ in our experiments) populations of $N$ randomly selected individuals (Line~\ref{algo:search:start}) and we select the population $P_0$ as the one with the individual having the lowest $F_{accuracy}$ (Line~\ref{algo:search:select:start}). \MAJOR{R3.1}{This choice is based on empirical observations from prior studies~\cite{Hazem:SEDE}, which demonstrated that using multiple seeds helps to reduce bias in the optimization process and ensures a broader exploration of the solution space. For instance, using five seeds strikes a balance between computational efficiency and the reliability of search results in related search-based testing tasks.}
All the individuals in $P_0$ are then used to update the archive based on Algorithm~\ref{alg:update} (Line~\ref{algo:search:firstUpdate}). In practice, the archive is populated with individuals whose distance is above $T_{diversity}$ and lead to worst accuracy.
 
As per NSGA-II, the population undergoes a non-dominated sorting process and crowding distance calculation (Lines \ref{algo:search:sort}-\ref{algo:search:crowd}).

The algorithm's core is a loop that iterates until the generation count reaches the specified maximum $g_{max}$. Within each generation $g$, we use NSGA-II's tournament selection method to select parent individuals for the next generation (Line~\ref{algo:search:parents}). 
Offspring generation follows, employing polynomial mutation and simulated binary crossover (Line~\ref{algo:search:offspring}).
Then, we compute $F_{accuracy}$ for each individual in $Q_g$ (Line~\ref{algo:search:compute:metrics}). 
We also compute $F_{similarity}$ for each individual in $R_g$, with $R_g$ being the union of the current population $P_g$ with the offspring population $Q_g$ (Line~\ref{algo:search:loop:diversity}). Recomputing $F_{similarity}$ for $P_g$ is necessary because, in the previous iteration, new individuals have been added to the archive and, therefore, distances have changed.
We then try to add each image to the archive based on Algorithm \ref{alg:update} (Line \ref{algo:search:loop:update}). Finally, since the archive changed, we update $F_{similarity}$ for the individuals in $R_g$ (Line~\ref{algo:search:loop:update_diversity}). 
$R_g$ is then ranked using the non-dominated sorting process (Line \ref{algo:search:loop:rank}) and the top individuals, based on rank and crowding distance, form the new population $P_{t+1}$ for the subsequent generation (Lines \ref{algo:search:loop:crowd}-\ref{algo:search:loop:select}).

The algorithm concludes by returning the populated archive, which includes diverse individuals leading to minimal DNN accuracy.

%% file: evaluation.tex
\section{Empirical Evaluation}
\label{sec:evaluation}

We report on an empirical assessment of \APPR aiming to address the following research questions:

\textbf{RQ1}: How does \APPR compare to alternative DNN testing solutions in terms of test effectiveness? Alternatives include approaches generating test images by relying on GANs (but not simulators), random baselines, and alternative implementations of our approach (ablation study). We discuss what approach leads to inputs leading to the lowest DNN accuracy and having the highest diversity.

\textbf{RQ2}: How does \APPR compare to alternative DNN testing solutions to improve the DNN? Images generated for DNN testing can also be used to retrain a DNN; we therefore aim to determine what approach generates images that, when used for DNN retraining, increase DNN accuracy the most.


\subsection{Subjects of the Study}
\MAJORBEGIN{}
As subjects of our study, we consider two distinct subjects: 
1) The DeeplabV3 DNN~\cite{chen2017deeplab} from the Gluoncv library~\cite{gluoncvnlp2020}, which is a state-of-the-art semantic segmentation DNN that classifies each pixel image into a predefined class. This DNN focuses on urban environments relying on the Cityscapes dataset~\cite{sagar2021semantic}. 
2) A DeeplabV3 DNN trained for object detection on Mars, utilizing the AI4MARS dataset~\cite{swan2021ai4mars}. This DNN implements a key task for the navigation of Mars rovers; specifically, it detects objects such as rocks, sand, and bedrocks.
This subject was developed in the context of a project with the European Space Agency.

For the urban environment case, DeeplabV3 was initially trained by its developers on the Cityscapes dataset, which contains a variety of urban street scenes from different European cities. The dataset consists of a variety of images captured at a resolution of 1024 by 2048 pixels in the streets of 50 different European cities, primarily in Germany. Each real-world image comes with a segmentation mask representing its ground truth. The DeeplabV3 training set consists of 2975 images; however, its test set comes without ground truth and, therefore, for testing, we rely on the Cityscapes validation set (500 images). \MAJOR{R3.3}{The average $IoU_{car}$ metric (see Section~\ref{sec:approach}) is particularly high: 71\% (min=00\%, first-quartile=62\%, third-quartile=91\%, max=97\%).}

For the Mars navigation subject, we trained the DNN on the AI4MARS dataset, which provides annotated images of the Martian surface. The simulator for generating synthetic Martian images was built using Unreal Engine~\cite{unrealengine} with the same plugins used by AirSim. The simulator was configured to simulate realistic Martian terrain and lighting conditions. The Airsim plugins are added to: simulate Vehicle dynamics (including interaction with sand and soil), add LIDAR for object detection, provide Python API for real-time control of vehicle, enabling communication between our search algorithm and the simulated vehicle. For segmentation on the AI4MARS dataset, DeepLabV3 achieved an average mIoU of 49\% (min=0\%, first-quartile=1\%, third-quartile=77\%, max=99\%).
\MAJOREND{}



\subsection{Assessed techniques}
\label{sec:empirical:techniques}
In our experiments, we assess a \APPR prototype, that relies on \MAJOR{R2.1}{ the original pix2pixHD GAN~\cite{pix2pixHD} and AirSim 1.8.1~\cite{airsim1.8.1} for the urban case study, and Unreal Engine 5 for the Martian case study}. We also extended the NSGA-II algorithm provided by the Pymoo library~\cite{pymoo}.

To address our RQs, we compare \APPR with a random baseline, two variants of \APPR (for the purpose of performing ablation studies), \MAJOR{R2.1}{ TACTIC~\cite{li2021testing}, and DeepJanus~\cite{riccio2020model}. We selected DeepJanus because it may lead to better results if a DNN tends to fail where model behaviour changes. 
For DeepJanus, in addition to the original implementation provided by its authors, we developed an extended version that integrates a GAN, as described below.}

Our random baseline relies only on the GAN-based input generation component; specifically, it generates realistic images from simulator segmentation maps generated using randomly selected values for the simulator parameters (e.g., position of the car in Airsim). For brevity, we refer to the images from our random baseline as \emph{random realistic images}.
We generate as many random images as \APPR (i.e., $12*100=1200$).

We defined three additional \APPR variants (hereafter, \APPRMONO, \APPRPIXEL, and \APPRNOGAN) to assess how $F_{similarity}$ and the feature-based distance contribute to \APPR results. Specifically, \APPRMONO employs a single objective search where our algorithm is configured with $F_{accuracy}$ as the sole fitness function and all the images are added to the archive. \APPRPIXEL relies on the same algorithms as \APPR but, instead of relying on  feature-based distance, it computes the distance between two realistic images as the percentage of pixels that do not match. Such distance metric matches the pixel accuracy metric, which is commonly used to assess semantic segmentation DNNs. \APPRNOGAN uses the same process as \APPR, but without the GAN component, therefore, both fitnesses are computed using the simulator images.


\input{tables/config}

Table~\ref{tab:parameters} provides the hyper-parameters used for \APPR, \APPRMONO, \APPRPIXEL, and \APPRNOGAN. For population size and number of generations, we selected the same values adopted by related search-based approaches with similar objectives (i.e., minimize accuracy and maximize diversity)~\cite{riccio2020model}. We selected the mutation and crossover probabilities that proved to be effective in our previous work on simulator-based image generation~\cite{Hazem:SEDE}.
Thresholds were determined based on 1000 randomly generated images, as described in previous sections. The two different $T_{similarity}$ thresholds refer to the two different distance metrics used in \APPR and \APPRPIXEL.

TACTIC is a state-of-the-art DNN testing approach that leverages a MUNIT GAN trained to alter input images by changing their weather conditions (night, sunshine, rain, snow); further, it integrates a search-based approach which alters the latent space vector in such a way that the generated images lead to the worst accuracy for the DNN under test. To apply TACTIC, we randomly select 220 images from the Cityscapes validation set and apply the five weather conditions supported by TACTIC (night, rain, snow night, snow daytime, and sunny conditions), each leading to a distinct new test image. This process generates a total of 1100 images. \MAJOR{R3.6}{Since TACTIC was originally trained to generate driving conditions, we consider it for the Urban environment only; indeed, it makes little sense to add snow and rain effects on Mars images, as well as, changing illumination to make it similar to Earth.}

\MAJOR{R2.1}{DeepJanus is multi-objective search-based approach that searches for the boundary of the input space that is properly processed by the DNN; specifically, it looks for pair of images that are similar but leads to different DNN results, one of which is erroneous.}
\MAJOR{R2.1}{We integrate DeepJanus with our simulators with the same configurations as in the original paper. Furher, since DeepJanus exclusively produces simulated images, we explore the impact of using a GAN to generate more realistic images (hereafter, \DEEPJANUSGAN). To achieve this, we incorporated the Pix2pixHD model, as utilized in \APPR, into the DeepJanus workflow to transform simulated images into realistic ones; precisely, \DEEPJANUSGAN, during search, still generates simulator images and ground truth as in \DEEPJANUS, but, instead of providing a simulator image to the DNN under test, it first generates a realistic image using Pix2pixHD and then feeds it to the DNN under test.}


\subsection{RQ1: Effectiveness}
\label{sec:evaluation:RQ1}
\label{sec:evaluation:RQ1_1}

\subsubsection{Design and measurements}

\MAJOR{R3.6}{For each assessed technique, we compute the $IoU_{car}$, for the urban case study, and $mIoU$ for the Martian case study (see Section~\ref{approach:accuracyFitness} for definitions),} obtained by the DeepLabV3 model with the generated images. Since we aim to discover critical DNN limitations, the most effective technique is the one leading to the lowest median, minimum, and first quartile.



However, since we also wish to obtain a set of diverse unsafe situations, we also assess  diversity across the generated images by relying on the two distance metrics employed by \APPR: feature-based distance and pixel-based distance. Specifically, we compute the distance between every possible pair of images generated by the selected techniques and compare their distributions. An approach with higher diversity values is preferable to identify the different situations under which segmentations are erroneous.

To cope with the non-determinism characterizing all the considered techniques, we apply each of the selected techniques ten times.

We discuss the significance of the differences by relying on a non-parametric Mann-Whitney U-test (the U statistics is computed considering all the datapoints generated by the ten executions). Further, we address effect size by relying Vargha and Delaney's $\hat{A}_{12}$. The $A_{12}$ statistic, given observations (e.g., IoU) obtained with two techniques X and Y, indicates the probability that technique X leads to higher values than technique Y. Based on $A_{12}$, effect size is considered small when $0.56 \le A_{12} < 0.64$, medium when $0.64 \le A_{12} < 0.71$, and large when $A_{12} \ge 0.71$. Otherwise the two populations are considered equivalent~\cite{VDA}.
In contrast, when $A_{12}$ is below $0.50$, it is more likely that treatment X leads to lower values than treatment Y; effect size is small when $0.36 < A_{12} \le 0.44$, medium when $0.29 < A_{12} \le 0.36$, and large when $A_{12} \le 0.29$.

\subsubsection{Results} 

\emph{\textbf{Accuracy.}} \MAJOR{R3.6}{Table~\ref{tab:rq1_quality} and~\ref{tab:rq1_quality_mars} provide, for each of the selected approaches,  details on the performance of the subject DNNs for the urban and the Martian environment, respectively.} Column \emph{images} provides the average number of test images selected over the ten executions; for \APPR and its variants we report the images in the archive, for Random and TACTIC we report all the generated images, excluding, for the former, an average of 15  and 169 images, for the urban and Martin environments, respectively, that do not satisfy `${0}<\mathit{pixelsProportion}<{0.4}$'. The other columns provide descriptive statistics for $IoU_{car}$ and \MAJOR{R3.6}{$mIoU$}. Table \ref{tab:rq1_diversity_stats} and Table~\ref{tab:rq1_quality_stats_mars} report differences between each pair of techniques by providing p-values and $A_{12}$ for the urban and martian environments, respectively. We discuss the results observed with the urban environment first.

\APPRMONO is the approach leading to the largest number of images with low $IoU_{car}$ values; indeed, it shows the lowest average (\$0.48\$) and median (\$0.67\$) values. However, although their difference is always statistically significant (p-value $< 0.05$, Table \ref{tab:rq1_diversity_stats}), based on $A_{12}$, \APPRMONO, \APPR, \APPRPIXEL, and \DEEPJANUSGAN perform similarly and do not lead to practically significant differences.

Except for \APPRNOGAN, all the \APPR variants perform significantly better (lower $IoU_{car}$) than Random, and TACTIC (Table \ref{tab:rq1_diversity_stats}), which indicates that combining simulators driven by search with GAN is necessary for effective testing. For example, \APPR leads to a median $IoU_{car}$ of $0.77$ compared to $0.85$ and $0.82$ for Random and TACTIC, respectively. Further, the first quartile of all the GAN-based \APPR variants is zero, indicating that 25\% of the generated images do not correctly identify a single pixel belonging to cars. For Random, less than 25\% of the images lead to $IoU_{car}=0$ (first quartile is $0.58$), while TACTIC tends to lead to a higher $IoU_{car}$ ($\text{average}\ \mathit{IoU}_{car} = 0.81$), with $IoU_{car}$ never being equal to 0. Also, all the \APPR variants fare better than \DEEPJANUSGAN, thus showing that an appropriate search strategy is necessary to identify test inputs leading to failures. As for \DEEPJANUS, its $IoU$ is similar in terms of median to that of \APPRMONO and worse in terms of average.

Also, Table \ref{tab:rq1_quality_stats} shows that the difference between \APPR and both TACTIC and Random is practically significant, according to $A_{12}$.
\APPRNOGAN, instead, leads to median results ($0.84$) that are in-between those of TACTIC and Random, with $A_{12}$ indicating that \APPRNOGAN and TACTIC perform similarly, while Random performs slightly better than \APPRNOGAN ($A_{12}=0.43$). These results suggest that relying on simulator images only is not sufficient to drive the search towards cases where the DNN performs worse.

\begin{figure}
    \centering
    \includegraphics[width=\linewidth]{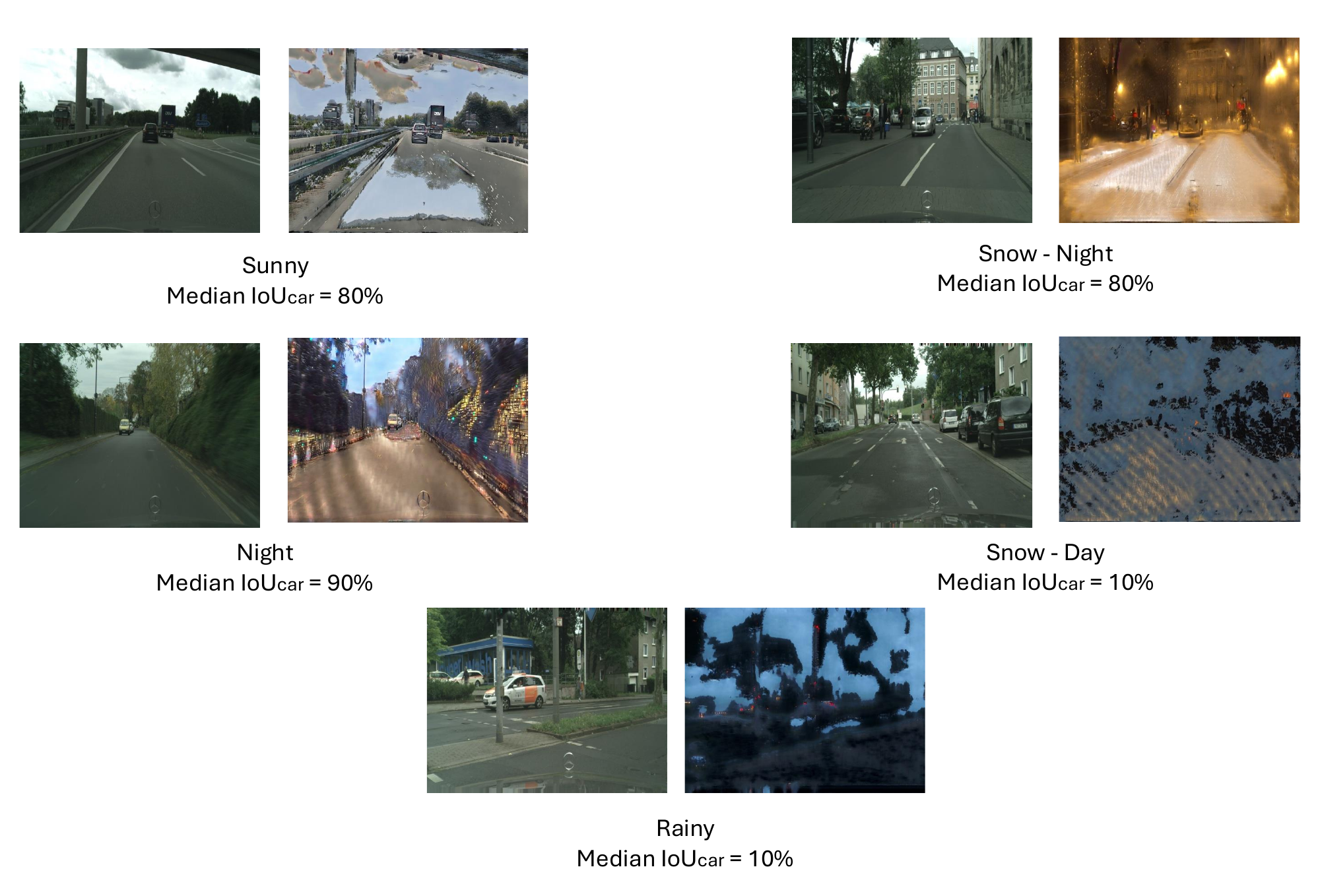}
    \caption{Examples of images generated by TACTIC for various weather conditions (right), along with the original cityscapes image (left) and the median $IoU_{car}$ of the generated images for each respective weather condition.}
    \label{fig:example_tactic}
\end{figure}

\MAJOR{3.4}{
Concerning TACTIC, we further checked how the different weather conditions (rain, snow day, sunny, and night) perform. Figure~\ref{fig:example_tactic} illustrates an example of a generated image for each weather condition, alongside the median $IoU_{car}$ of the generated images for the respective conditions.}
\MAJORBEGIN{}
Our observations indicate that three conditions --- Sunny (median $IoU_{car}$ = 80\%, min = 50\%, max = 97\%, average = 78\%), Night (median $IoU_{car}$ = 90\%, min = 50\%, max = 94\%, average = 80\%), and Snow-Night (median $IoU_{car}$ = 80\%, min = 50\%, max = 91\%, average = 80\%) --- yield high median $IoU_{car}$ values, with the generated images appearing realistic. Conversely, the Rainy (median $IoU_{car}$ = 10\%, min = 1\%, max = 32\%, average = 10\%) and Snow-Day (median $IoU_{car}$ = 10\%, min = 2\%, max = 36\%, average = 10\%) conditions result in lower median $IoU_{car}$ values, with the generated images being unrealistic, with no observable cars, and therefore not matching the ground truth. Such images, as illustrated in Figure~\ref{fig:example_tactic}, are not useful for testing purposes. 
\MAJOREND{}

The slightly better performance of \APPRMONO can be explained by the fact that \APPRMONO, being single-objective, may be stuck in a particular region of the solution space, leading to outputs that are less diverse but all leading to minimal $IoU_{car}$. In contrast, \APPR adopts a multi-objective strategy, aiming to balance both low $IoU_{car}$ and diversity, which may lead to a higher $IoU_{car}$ (e.g., there might be regions of the solution space where the DNN performs poorly but not as bad as in the only region detected by \APPRMONO).

The random baseline appears to be less effective, with a high median $IoU_{car}$ of 85\%, which shows that the DNN under test performs generally well in the sampled input space and a search-based approach is needed to comprehensively uncover its limitations.

\MAJOR{R2.1}{\DEEPJANUS yields slightly better accuracy compared to \APPR and \APPRPIXEL with $A_{12}$ suggesting that \DEEPJANUS, \APPR, and \APPRPIXEL perform similarly. However, \APPRMONO outperforms \DEEPJANUS, which suggests that the single-objective search used by \APPRMONO leads to better results than the one used by \DEEPJANUS. 
Further, after visual inspection, we realized that most the failures triggered by \DEEPJANUS are due to borderline cases with cars appearing far away and occupying few pixels in the image. Though such sceneries can be generated by simulators, they are likely to be irrelevant in practice as the position of cars that are far away are unlikely to affect the behaviour of the ego vehicle and thus be relevant for testing. 
More precisely, because \DEEPJANUS looks for frontier behaviour, with pairs of images that are very similar but where one of the two leads to failures, it tends to generate images where cars are far from the ego vehicle.} 

\MAJORBEGIN{}
The higher accuracy of \DEEPJANUSGAN, which performs worse and significantly different from \APPR, is likely due to the fact that, when applying GANs, far away cars disappear from the picture. Indeed, the training set does not contain cars occupying only few pixels, because 
such far-away cars are irrelevant for a segmentation DNN in an automotive system and thus not important for testing purposes.
\MAJOREND{}

\MAJOR{R3.6}{For the Martian environment, \APPR achieved a median $mIoU$ of 0.49, outperforming \APPRMONO (0.54), \APPRPIXEL (0.52), Random (0.64), and \DEEPJANUSGAN (0.60). Similar to the urban environment, the $A_{12}$ values in Table~\ref{tab:rq1_quality_stats_mars} indicate that \APPRMONO, \APPR, and \APPRPIXEL perform similarly, although their differences are statistically significant ($p < 0.05$).}  

\MAJOR{R3.7, R2.5}{
The key difference between the urban and Martian environments is that the lowest accuracy values are observed with \APPRNOGAN and \DEEPJANUS: 0.32 and 0.48 in the latter, as opposed to \APPRMONO in the former. We attribute this result to \APPRNOGAN and \DEEPJANUS relying on simulated images that are often unrealistic, thereby increasing the likelihood of mispredictions. This can be explained by the low fidelity of the Mars simulator when compared to AirSim, as depicted in Fig.~\ref{fig:example_mars}. To conclude, as reported for the urban environment, the use of GAN prevents the identification of failures that will not occur in real usage settings. In other words, when not using a GAN and relying on a low-fidelity simulator, many unrealistic and therefore irrelevant test inputs are generated.}


\input{tables/RQ1_quality}
\input{tables/RQ1_quality_stats}

\input{tables/RQ1_quality_mars}
\input{tables/RQ1_quality_stats_mars}




\input{tables/RQ1_diversity}
\input{tables/RQ1_diversity_stats}

\input{tables/RQ1_diversity_mars}
\input{tables/RQ1_diversity_stats_mars}
\emph{\textbf{Diversity.}} Table~\ref{tab:rq1_diversity} and Table~\ref{tab:rq1_diversity_mars} provide descriptive statistics for the diversity metrics collected for the selected techniques for the Urban and Martian environments, respectively. Additionally, Table~\ref{tab:rq1_diversity_stats} and Table~\ref{tab:rq1_diversity_stats_mars} provide p-values and $\hat{A}_{12}$ resulting from the comparisons of diversity across techniques for the Urban and Martian environments, respectively.
We discuss the urban environment first.

Considering feature-based distance, \APPR stands out with the highest median (14.97) and average (15.25) values. This result is likely due to \APPR integrating feature-based distance in $F_{similarity}$. Further, \APPR performs significantly better than other \APPR variants ($p-value < 0.01$) and has a significantly higher probability of performing better than other \APPR variants (medium effect size, based on $\hat{A}_{12}$). 
Expectedly, the lack of a diversity objective in \APPRMONO leads to images that are less diverse than those of \APPR. The lower diversity of \APPRPIXEL compared to \APPR confirms that a fitness based on pixel distance is not effective to obtain feature-based diversity. Last, the lower median obtained by \APPRNOGAN ($12.90$) is likely due to the lower fidelity of simulator images compared to the ones generated by the GAN (e.g., simulator images lack details, leading to images that are more similar to each other).

\APPR performs significantly better, with high effect size, than TACTIC and Random. TACTIC has a feature-based distance that is similar across pairs of images, between $8.09$ and $10.49$ (avg. is $10.06$). Since in our experiments TACTIC works by modifying a set of 220 randomly selected images, and since these images present a feature-based distance of at least $8.09$, it seems that the degree of diversity introduced by TACTIC through changes in weather conditions is limited. \APPR, instead is more effective in creating diverse images as 75\% (1st quartile) of the pairs of images generated by \APPR have a distance higher than $12.29$, which is higher than the best results achieved by TACTIC. Further 25\% of \APPR images yield a distance higher than $18.08$.

The random baseline leads to images with low feature-based distance, with a median of $10.99$ and a maximum of $12.34$; it shows that the mere sampling of the input space does not lead to identifying diverse images and that search guidance is needed.

\MAJOR{R2.1}{\APPR significantly outperforms \DEEPJANUS and \DEEPJANUSGAN. Their lower diversity suggests that their search algorithm, which focuses on identifying the boundaries of correctly processed inputs, is unexpectedly less effective than NSGA-II in exploring the search landscape. 
\DEEPJANUSGAN achieves higher diversity values ($29.71$) than \DEEPJANUS  thus confirming that GANs, by avoiding unrealistic cases (e.g., cars with few visible pixels, as discussed above), prevents the search from getting stuck in unrealistic, similar images.}

Considering pixel-based distance, all the \APPR variants have the same average distance but \APPR still performs better than the others, with a small to high effect size. Such results confirm our observations above for \APPRMONO and \APPRNOGAN. Further, the comparison with \APPRPIXEL indicates that a fitness driven by pixel-based distance is less effective than a feature-based distance fitness to maximize pixel-based diversity across images. 
This indicates that, for the search algorithm, it is easier to generate pixel-wise diverse images when driven by a feature-based fitness. Note that  augmenting feature-based distance implies changing the elements depicted in the images and, consequently, the pixel distance. This result may thus be explained by the fact that most of the changes in simulator inputs lead to pixel-based distances that are significant (i.e., above $T_{diversity}$), whereas only pixel changes that have a strong impact on the elements in the images can result into a significant feature-based distance. Consequently, feature-based distance simplifies the task of the search algorithm to explore the input space towards the search objectives. 
Also, \APPR performs significantly better than TACTIC and Random, with small and high effect size, respectively. As per feature-based distance, Random presents lower diversity than \APPR (ranges are [0.3,0.76] for Random vs. [0.14,1.00] for \APPR), thus confirming that random sampling is not sufficient to identify the unsafe parts of the input space. TACTIC's third quartile shows that only 25\% of TACTIC's image pairs have a pixel-based distance higher than 0.68 (i.e., 68\% of the pixels differ) while 95\% (5th percentile) of the image pairs generated by \APPR present a distance higher than 0.94 (almost all the pixels differ). \MAJOR{2.1}{Last, the pixel-based distance of \APPR images is significantly larger than that of \DEEPJANUS and \DEEPJANUSGAN images, with medium to large effect sizes, thus confirming the differences discussed above.}

\MAJOR{R3.6}{In the Martian environment, \APPR achieved a median diversity of 14.75 for feature-based distance that is significantly higher than \APPRMONO (10.59), \APPRPIXEL (11.92), the random baseline (11.36), \DEEPJANUS (12.81), and \DEEPJANUSGAN (9.52). However, \APPRNOGAN outperformed \APPR in terms of feature-based diversity, which we attribute to \APPRNOGAN relying on the lower-fidelity outputs of the simulator; indeed, such simulator images are likely to cause erroneous outputs and therefore \APPRNOGAN is likely to generate images that are spread throughout the whole input space, thus leading to larger diversity. In all likelihood, this phenomenon is not observed with \DEEPJANUS because, by generating images on the boundaries of the input space, it likely leads to more similar images (e.g., images containing roads where cars can be positioned far away).} 

\MAJOR{R3.6}{For pixel-based diversity, \APPR outperformed all the other approaches with a median pixel distance of 1.00. Instead, with feature-based distance, \APPRNOGAN outperformed \APPR, such a difference is likely due to the nature of the two distance metrics; indeed, the pixel distance may capture dissimilarities that are practically irrelevant. For example, two images may have slightly different luminosity, thus leading to high pixel-based distance, but contain the same objects thus yielding low feature-based distance.}


  
To summarize, \APPRMONO fares better than \APPR and \APPRPIXEL in terms of accuracy. However, \APPR yields more diversity (both in terms of features and pixels) than \APPRMONO and \APPRPIXEL. In other words, \APPR and \APPRMONO represent two different trade-offs and which one to adopt depends on whether accuracy or diversity is prioritized. Since \APPRMONO focuses solely on minimizing DNN accuracy, without explicitly considering diversity, it is highly effective at generating images that cause DNN failures, but at the cost of producing less diverse inputs. We should note, however, that \APPR is nevertheless able to detect a large number of cases where the DNN predicts poorly, as demonstrated by the zero accuracy cases representing more that 25\% of generated inputs (first quartile of the $IoU_{car}$ distribution in Table~\ref{tab:rq1_quality}). Last, there is no particular reason to rely on \APPRPIXEL as it provides no advantages in terms of diversity or accuracy. With respect to other alternative techniques, as presented above, they tend to either not be effective at triggering failures or generate many unrealistic and irrelevant test inputs.

\subsection{RQ2: Retraining}

\subsubsection{Design and measurements}

We use the images generated by the approaches considered for RQ1 to retrain our case-study subject DNN. Further, to account for non-determinism, we repeat the process 10 times. For each approach and each retraining iteration, we use the images generated in one run of the approach to augment the Cityscapes training set (2975 images), \MAJOR{R3.6}{for the urban environment, and the AI4MARS training set (16064 images), for the Martian environment}. We then retrain the DNN using the augmented dataset, maintaining the same configuration settings (100 epochs) as those used by the DeeplabV3 developers~\cite{chen2017rethinking}. To avoid introducing bias due to a different number of images in the augmented datasets, for each approach, we randomly select up to 900 images from the set of images generated in RQ1. In addition to the approaches presented in RQ1 (TACTIC, random baseline, \DEEPJANUS, and \DEEPJANUSGAN), we compare \APPR to the training of the DeeplabV3 model for 100 additional epochs using the whole original training set, to determine the degree of improvement achievable with a longer training time, without generating additional training images. 

We measure the $IoU_{car}$ obtained by all the retrained DNNs on the Cityscapes validation set, which enables us to assess if retraining the DNN with test images improves its performance with real-world inputs, the target of the DNN under test. We analyze on the $IoU_{car}$ distribution and, as per RQ1, we discuss significance of the differences and effect sizes by relying on the results of Mann-Whitney U-test and Vargha and Delaney's $A{12}$, respectively.

For the Martian environment, we measure the $mIoU$ obtained by all the retrained DNNs on the AI4MARS test set (966 images). 

\subsubsection{Results}
\MAJOR{R3.6}{Tables~\ref{tab:rq2_results} and~\ref{tab:rq2_results_mars} provide descriptive statistics for $IoU_{car}$ and $mIoU$ from testing the retrained models in the urban and Martian environments, respectively. These tables summarize the performance improvements achieved through retraining with outputs generated by different approaches. Tables~\ref{tab:rq2_stats} and~\ref{tab:rq2_stats_mars} present the p-values and $\hat{A}_{12}$ values from statistical comparisons of $IoU_{car}$ and $mIoU$ across retrained models, capturing the significance and effect sizes of the observed differences.}

For the urban environment, the DNN retrained with the output of \APPR has the highest median performance at $0.90$, followed by \DEEPJANUS ($0.87$), \APPRMONO ($0.86$), \DEEPJANUSGAN ($0.86$), \APPRNOGAN ($0.85$), and \APPRPIXEL ($0.85$). 
\APPR not only improves ($+0.08$) over the original pre-trained DeeplabV3 model (median $IoU_{car}=0.82$), but fare much better than the DeeplabV3 retrained with the cityscapes training set ($0.84$, $+0.06$), a set of random realistic images ($0.85$, $+0.05$),  TACTIC ($0.84$, $+0.06$), \DEEPJANUS ($0.87$, $+0.05$) and \DEEPJANUSGAN ($0.84$, $+0.04$) outputs.

The difference between \APPR and its variants is statistically ($\mathit{p-value} < 0.05$) and practically ($\hat{A}_{12} \ge 0.56$) significant, which indicates that integrating a diversity objective and feature distance as fitness functions leads to better results. 

\MAJOR{R3.6}{Also the difference between \APPR and the two \DEEPJANUS variants is statistically and practically significant. The fact that \APPR outperforms \DEEPJANUSGAN indicates that focusing on corner cases (i.e., the frontier of behaviours found by \DEEPJANUSGAN) is not effective to retrain a DNN. 
Further, the likelihood of \DEEPJANUS and \DEEPJANUSGAN performing better than random is very low ($0.55$ and $0.52$).}

In general, we can conclude that relying on simulator images only (i.e., \APPRNOGAN and \DEEPJANUS) leads to worse results compared to retraining with GAN-generated images, thus justifying our choice.

\MAJOR{R3.6}{Further, models retrained relying on search to drive the generation of scenarios (\APPR and \DEEPJANUS variants) outperform the other approaches.}  Such results suggest that relying on changes in weather conditions (i.e., TACTIC) limits retraining and does not yield better results than  what is achievable with additional training epochs. 
Further, TACTIC leads to results that do not significantly differ from the ones obtained with the original DeeplabV3 model, with a model retrained with additional epochs, or with a set of random realistic images ($0.46 < A_{12} <0.54$).

\MAJOR{R3.6}{For the Martian environment, \APPR again demonstrated superior performance, achieving the highest median $mIoU$ (0.53, +0.09 over the original model). These results highlight the effectiveness of \APPR in generating both failure-inducing and diverse inputs, which, when used for retraining, improves DNN robustness in challenging environments. The difference between \APPR and its variants is statistically ($\mathit{p-value} < 0.05$) and practically ($\hat{A}_{12} \ge 0.56$) significant, thus confirming the effectiveness of integrating a diversity objective and feature distance as fitness functions.}


\MAJORBEGIN{}

Like in the urban environment, \APPR ($mIoU=0.53$) outperforms \DEEPJANUS ($0.50$) and 
\DEEPJANUSGAN ($0.50$), thus confirming to be a better choice for retraining.
Also, \APPR outperforms a random approach ($mIoU=0.49, +0.05$), with a practically significant difference ($A_{1,2}=0.61$), while \DEEPJANUS and \DEEPJANUSGAN perform similar to random ($A_{1,2}=0.50$).

\MAJOREND{}

Concluding, our results show that \APPR leads to the best retraining results, thus demonstrating that (1) simulator images are not as effective as realistic images to retrain DNNs, \MAJOR{R3.6}{(2) relying on images belonging to the frontier of DNN behaviours is not the best approach for DNN retraining, and} (3) relying on GANs to generate images that simply change weather conditions is not as effective as combining GAN with a simulator to generate realistic images for DNN retraining.

\input{tables/RQ2_results}

\input{tables/RQ2_stats}

\input{tables/RQ2_results_mars}
\input{tables/RQ2_stats_mars}

\subsection{Threats to validity}

To address \emph{internal validity} threats, we carefully tested our implementation of \APPR and its variants, and further relied on the original version of TACTIC and \DEEPJANUS provided by their authors.
\MAJOR{R3.7}{Still related to internal validity, to prevent the generation of unrealistic images with GANs, our approach integrates a strategy to eliminate images that are likely to be unrealistic (i.e., images where the DNN performs better with the simulated image). Such heuristic seems effective since RQ2 results show that retraining the DNN with \APPR images leads to improved results with a real-world test set. If \APPR images were not realistic, the performance of the DNN would have shown limited or no improvement because it would have learned from images that are unrealistic, which is what happens with approaches without GAN. However, it is still possible that our filter may not be fully effective. In the future, we aim to extend our approach with additional methods such as histogram analysis~\cite{HIST} and other state-of-the-art approaches~\cite{TonellaAmbiguity,RiccioOOD} for the identification of out-of-domain and out-of-distribution images.}

To address \emph{conclusion validity} threats, we repeated our experiments 10 times and relied on non-parametric statistical analysis (Mann Whitney U-Test) and effect size (Vargha and Delaney's) tests.

\MAJOR{R3.6}{To address \emph{generalizability threats}, we considered two instances of a state-of-the-art DNN implementing a computer vision task that is key for safety-critical systems (i.e., image segmentation), for two very different environments (urban and Martian). We therefore expect our results to be applicable across many industrial contexts, from automotive to robotics.}



%% file: tables/config.tex
\begin{table}[t]
\centering
\caption{Configuration of \APPR}
\label{tab:parameters}
\begin{tabular}{lcc}
\hline
\textbf{Parameter} & \textbf{Value} \\
\hline
Population size & 12 \\
Generations & 100 \\
Mutation probability & 0.3 \\
Crossover probability & 0.7 \\
$T_{similarity}$ (for \APPR) &  18 \\
$T_{similarity}$ (for \APPRPIXEL) &  0.02 \\
$T_{relevance}$ & 0.1 \\
\hline
\end{tabular}
\end{table}

%% file: tables/RQ1_quality.tex
\begin{table*}[t]
\smaller
\centering
\caption{RQ1-Accuracy assessment. Descriptive statistics for $IoU_{car}$ obtained with the generated images for the Urban environment.}
\label{tab:rq1_quality}
\begin{tabular}{|l|l|lllllll|}
\hline
\textbf{} & \multicolumn{1}{c|}{\textbf{\#images}} & \multicolumn{7}{c|}{$\mathbf{IoU_{car}} \textbf{distribution}$} \\ \hline
\textbf{} & \textbf{}& \multicolumn{1}{c|}{\textbf{min}} & \multicolumn{1}{c|}{\textbf{max}} & \multicolumn{1}{c|}{\textbf{median}} & \multicolumn{1}{l|}{\textbf{5th percentile}} & \multicolumn{1}{l|}{\textbf{1st quartile}} & \multicolumn{1}{l|}{\textbf{3rd quartile}} & \textbf{average} \\ \hline
\textbf{\APPRMONO}  & 893& \multicolumn{1}{l|}{\textbf{0.00}}& \multicolumn{1}{l|}{0.99}& \multicolumn{1}{l|}{\textbf{0.67}}& \multicolumn{1}{l|}{\textbf{0.00}}  & \multicolumn{1}{l|}{\textbf{0.00}}& \multicolumn{1}{l|}{0.89}& \textbf{0.48} \\ \hline
\textbf{\APPR} & 721& \multicolumn{1}{l|}{\textbf{0.00}}& \multicolumn{1}{l|}{0.99}& \multicolumn{1}{l|}{0.77}& \multicolumn{1}{l|}{\textbf{0.00}}  & \multicolumn{1}{l|}{\textbf{0.00}}& \multicolumn{1}{l|}{0.91}& 0.54 \\ \hline
\textbf{\APPRPIXEL}& 783& \multicolumn{1}{l|}{\textbf{0.00}}& \multicolumn{1}{l|}{0.99}& \multicolumn{1}{l|}{0.76}& \multicolumn{1}{l|}{\textbf{0.00}}  & \multicolumn{1}{l|}{\textbf{0.00}}& \multicolumn{1}{l|}{0.92}& 0.52 \\ \hline

\textbf{\APPRNOGAN} & 1190  & \multicolumn{1}{l|}{\textbf{0.00}}& \multicolumn{1}{l|}{1.00}& \multicolumn{1}{l|}{0.84}& \multicolumn{1}{l|}{0.21}  & \multicolumn{1}{l|}{0.68}& \multicolumn{1}{l|}{1.00}& 0.78 \\ \hline

\textbf{Random} & 1185  & \multicolumn{1}{l|}{\textbf{0.00}}& \multicolumn{1}{l|}{1.00}& \multicolumn{1}{l|}{0.85}& \multicolumn{1}{l|}{\textbf{0.00}}  & \multicolumn{1}{l|}{0.58}& \multicolumn{1}{l|}{0.95}& 0.69 \\ \hline

\textbf{TACTIC} & 1100  & \multicolumn{1}{l|}{{0.02}}& \multicolumn{1}{l|}{\textbf{0.97}}& \multicolumn{1}{l|}{0.82}& \multicolumn{1}{l|}{0.69}  & \multicolumn{1}{l|}{0.75}& \multicolumn{1}{l|}{0.88}& 0.81 \\ \hline

\textbf{\DEEPJANUS} & 943  & \multicolumn{1}{l|}{\textbf{0.00}}& \multicolumn{1}{l|}{0.99}& \multicolumn{1}{l|}{0.74}& \multicolumn{1}{l|}{0.26}  & \multicolumn{1}{l|}{0.59}& \multicolumn{1}{l|}{\textbf{0.82}}& 0.68 \\ \hline

\textbf{\DEEPJANUSGAN} & 954  & \multicolumn{1}{l|}{0.01}& \multicolumn{1}{l|}{0.99}& \multicolumn{1}{l|}{0.87}& \multicolumn{1}{l|}{0.40}  & \multicolumn{1}{l|}{0.76}& \multicolumn{1}{l|}{0.94}& 0.81 \\ \hline

\end{tabular}%
\\
Note: best (i.e., lowest) results (per column) in bold.
\end{table*}

%% file: tables/RQ1_quality_stats.tex
\begin{table*}[t]
\tiny
\centering
\caption{RQ1-Accuracy assessment. p-values and $A_{12}$ for the data in Table~\ref{tab:rq1_quality}.}
\label{tab:rq1_quality_stats}
\begin{tabular}{|@{}l|@{}l|@{}l|@{}l|@{}l|@{}l|@{}l|@{}l|@{}l|@{}l|@{}l|@{}l|@{}l|@{}l|@{}l|@{}l|@{}l|}
\hline
 & \multicolumn{2}{l|}{\textbf{\APPRMONO}}& \multicolumn{2}{l|}{\textbf{\APPR}}  & \multicolumn{2}{l|}{\textbf{\APPRPIXEL}}& \multicolumn{2}{l|}{\textbf{\APPRNOGAN}} & \multicolumn{2}{l|}{\textbf{Random}}  & \multicolumn{2}{l|}{\textbf{TACTIC}} & \multicolumn{2}{l|}{\textbf{\DEEPJANUS}} & \multicolumn{2}{l|}{\textbf{\DEEPJANUSGAN}} \\ \hline
 &  {\textbf{\VDA{}}} & \textbf{p-value}&  {\textbf{\VDA{}}} & \textbf{p-value} &  {\textbf{\VDA{}}} & \textbf{p-value} &  {\textbf{\VDA{}}} & \textbf{p-value} &  {\textbf{\VDA{}}} & \textbf{p-value} &  {\textbf{\VDA{}}} & \textbf{p-value} &  {\textbf{\VDA{}}} & \textbf{p-value} &  {\textbf{\VDA{}}} & \textbf{p-value} \\ \hline

\textbf{\APPRMONO}  &  {} &  &  {0.45} & 5.00E-28 &  {0.47} & 8.00E-13 &  {\textbf{0.28}} & 0.00E+00 &  {\textbf{0.34}} & 3.00E-298 &  {\textbf{0.35}} & 1.00E-64 &  {0.47} & 2.19E-09&  {\textbf{0.33}} & 7.53E-301\\ \hline

\textbf{\APPR} &  {0.55} & 5.00E-28 &  &  &  {0.52} & 1.00E-03 &  {\textbf{0.33}} & 0.00E+00 &  {\textbf{0.39}} & 1.14E-139 &  {\textbf{0.41}} & 9.00E-21 &  {0.53} & 9.24E-09&  {\textbf{0.38}} & 1.36E-131\\ \hline

\textbf{\APPRPIXEL} &  {0.53} & 8.00E-13 &  {0.48} & 2.00E-03 &  &  &  {\textbf{0.31}} & 0.00E+00 &  {\textbf{0.38}} & 8.00E-161 &  {\textbf{0.41}} & 1.00E-22 &  {0.52} & 4.69E-03&  {\textbf{0.38}} & 2.41E-150\\ \hline

\textbf{\APPRNOGAN} &  {\textbf{0.72}} & 0.00E+00 &  {\textbf{0.67}} & 0.00E+00 &  {\textbf{0.69}} & 0.00E+00 &  &  &  {\textbf{0.57}} & 0.00E+00 &  {0.55} & 0.00E+00 &  {\textbf{0.73}} & 1.88E-133&  {\textbf{0.61}} & 1.19E-33\\ \hline

\textbf{Random} &  {\textbf{0.66}} & 3.00E-298 &  {\textbf{0.61}} & 1.14E-135 &  {\textbf{0.62}} & 8.00E-161 &  {\textbf{0.43}} & 0.00E+00 &  &  &  {\textbf{0.54}} & 5.10E-05 &  {\textbf{0.66}} & 2.23E-210&  {0.50} & 6.95E-01\\ \hline

\textbf{TACTIC} &  {\textbf{0.64}} & 1.00E-64 &  {\textbf{0.58}} & 9.00E-21 &  {\textbf{0.59}} & 1.00E-22 &  {0.45} & 0.00E+00 &  {0.46} & 5.10E-05 &  &  &  {\textbf{0.73}} & 3.40E-55&  {0.45} & 2.15E-04\\ \hline

\textbf{\DEEPJANUS} &  {0.53} & 2.19E-09&  {0.47} & 9.24E-09&  {0.48} & 4.69E-03&  {\textbf{0.27}} & 1.88E-133&  {\textbf{0.34}} & 2.23E-210&  {\textbf{0.27}} & 3.40E-55&  &  &  {\textbf{0.29}} & 2.18E-297\\ \hline

\textbf{\DEEPJANUSGAN} &  \textbf{0.67} & 7.53E-301&  {\textbf{0.62}} & 1.36E-131&  {\textbf{0.62}} & 2.41E-150&  {\textbf{0.39}} & 1.19E-33&  {0.50} & 6.95E-01&  {0.55} & 2.15E-04&  {\textbf{0.71}} & 2.18E-297&  &  \\ \hline

\end{tabular}%

\vspace{1mm}
Note: tangible (based on $A_{12}$) differences in bold; for each pair of approaches, the best approach is the one with  $A_{12} < 0.50$ and its name on the row. 
\end{table*}

%% file: tables/RQ1_quality_mars.tex
\begin{table*}[t]
\smaller
\centering
\caption{RQ1-Accuracy assessment. Descriptive statistics for $mIoU$ obtained with the generated images for the Martian environment.}
\label{tab:rq1_quality_mars}
\begin{tabular}{|l|l|lllllll|l|}
\hline
\textbf{} & \multicolumn{1}{c|}{\textbf{\#images}} & \multicolumn{7}{c|}{$\mathbf{mIoU}\_\textbf{distribution}$}  \\ \hline
\textbf{} & \textbf{} & \multicolumn{1}{c|}{\textbf{min}} & \multicolumn{1}{c|}{\textbf{max}} & \multicolumn{1}{c|}{\textbf{median}} & \multicolumn{1}{l|}{\textbf{5th percentile}} & \multicolumn{1}{l|}{\textbf{1st quartile}} & \multicolumn{1}{l|}{\textbf{3rd quartile}} & \multicolumn{1}{l|}{\textbf{average}} \\ \hline

\textbf{\APPRMONO}  & 802 & \multicolumn{1}{l|}{\textbf{0.01}} & \multicolumn{1}{l|}{0.96} & \multicolumn{1}{l|}{0.54} & \multicolumn{1}{l|}{0.09}  & \multicolumn{1}{l|}{0.35} & \multicolumn{1}{l|}{0.73} & 0.52 \\ \hline

\textbf{\APPR} & 764 & \multicolumn{1}{l|}{\textbf{0.01}} & \multicolumn{1}{l|}{0.96} & \multicolumn{1}{l|}{0.49} & \multicolumn{1}{l|}{0.09}  & \multicolumn{1}{l|}{0.33} & \multicolumn{1}{l|}{0.69} & 0.50 \\ \hline

\textbf{\APPRPIXEL} & 718 & \multicolumn{1}{l|}{\textbf{0.01}} & \multicolumn{1}{l|}{0.95} & \multicolumn{1}{l|}{0.52} & \multicolumn{1}{l|}{0.20}  & \multicolumn{1}{l|}{0.39} & \multicolumn{1}{l|}{0.70} & 0.53 \\ \hline

\textbf{\APPRNOGAN} & 1151 & \multicolumn{1}{l|}{\textbf{0.01}} & \multicolumn{1}{l|}{0.97} & \multicolumn{1}{l|}{\textbf{0.32}} & \multicolumn{1}{l|}{\textbf{0.03}}  & \multicolumn{1}{l|}{\textbf{0.12}} & \multicolumn{1}{l|}{0.70} & \textbf{0.39}\\ \hline

\textbf{Random} & 1031 & \multicolumn{1}{l|}{\textbf{0.01}} & \multicolumn{1}{l|}{0.96} & \multicolumn{1}{l|}{0.64} & \multicolumn{1}{l|}{0.24}  & \multicolumn{1}{l|}{0.44} & \multicolumn{1}{l|}{0.83} & 0.62 \\ \hline

\textbf{\DEEPJANUS} & 986 & \multicolumn{1}{l|}{\textbf{0.01}} & \multicolumn{1}{l|}{\textbf{0.93}} & \multicolumn{1}{l|}{0.48} & \multicolumn{1}{l|}{\textbf{0.03}}  & \multicolumn{1}{l|}{0.20} & \multicolumn{1}{l|}{\textbf{0.67}} & 0.44 \\ \hline

\textbf{\DEEPJANUSGAN} & 954 & \multicolumn{1}{l|}{\textbf{0.01}} & \multicolumn{1}{l|}{0.98} & \multicolumn{1}{l|}{0.60} & \multicolumn{1}{l|}{0.20}  & \multicolumn{1}{l|}{0.42} & \multicolumn{1}{l|}{0.85} & 0.61 \\ \hline
\end{tabular}%
 
\vspace{1mm}
Note: best (i.e., lowest) results (per column) in bold.
\end{table*}

%% file: tables/RQ1_quality_stats_mars.tex
\begin{table*}[t]
\tiny
\centering
\caption{RQ1-Accuracy assessment for the Martian case study. p-values and $A_{12}$ for the data in Table~\ref{tab:rq1_quality_mars}.}
\label{tab:rq1_quality_stats_mars}
\begin{tabular}{|l|ll|ll|ll|ll|ll|ll|ll|}
\hline
 & \multicolumn{2}{l|}{\textbf{\APPRMONO}} & \multicolumn{2}{l|}{\textbf{\APPR}} & \multicolumn{2}{l|}{\textbf{\APPRPIXEL}} & \multicolumn{2}{l|}{\textbf{\APPRNOGAN}} & \multicolumn{2}{l|}{\textbf{Random}} & \multicolumn{2}{l|}{\textbf{\DEEPJANUS}} & \multicolumn{2}{l|}{\textbf{\DEEPJANUSGAN}} \\ \hline
 & \multicolumn{1}{l|}{\textbf{\VDA{}}} & \textbf{p-value} & \multicolumn{1}{l|}{\textbf{\VDA{}}} & \textbf{p-value} & \multicolumn{1}{l|}{\textbf{\VDA{}}} & \textbf{p-value} & \multicolumn{1}{l|}{\textbf{\VDA{}}} & \textbf{p-value} & \multicolumn{1}{l|}{\textbf{\VDA{}}} & \textbf{p-value} & \multicolumn{1}{l|}{\textbf{\VDA{}}} & \textbf{p-value} & \multicolumn{1}{l|}{\textbf{\VDA{}}} & \textbf{p-value} \\ \hline

\textbf{\APPRMONO} & \multicolumn{1}{l|}{} &  & \multicolumn{1}{l|}{0.53} & 1.10E-11& \multicolumn{1}{l|}{0.50} & 4.41E-01& \multicolumn{1}{l|}{\textbf{0.63}} & 4.55E-197& \multicolumn{1}{l|}{\textbf{0.38}} & 3.11E-153& \multicolumn{1}{l|}{\textbf{0.59}} & 1.80E-81& \multicolumn{1}{l|}{\textbf{0.40}} & 1.45E-116\\ \hline

\textbf{\APPR} & \multicolumn{1}{l|}{0.47} & 1.10E-11& \multicolumn{1}{l|}{} &  & \multicolumn{1}{l|}{\textbf{0.46}} & 5.22E-16& \multicolumn{1}{l|}{\textbf{0.61}} & 2.20E-136& \multicolumn{1}{l|}{\textbf{0.36}} & 4.16E-228& \multicolumn{1}{l|}{\textbf{0.56}} & 2.89E-34& \multicolumn{1}{l|}{\textbf{0.37}} & 5.03E-178\\ \hline

\textbf{\APPRPIXEL} & \multicolumn{1}{l|}{0.50} & 0.440865115& \multicolumn{1}{l|}{0.54} & 5.22E-16& \multicolumn{1}{l|}{} &  & \multicolumn{1}{l|}{\textbf{0.65}} & 2.54E-239& \multicolumn{1}{l|}{\textbf{0.38}} & 6.00E-146& \multicolumn{1}{l|}{\textbf{0.59}} & 2.76E-85& \multicolumn{1}{l|}{\textbf{0.40}} & 1.90E-103\\ \hline

\textbf{\APPRNOGAN} & \multicolumn{1}{l|}{\textbf{0.37}} & 4.55E-197& \multicolumn{1}{l|}{\textbf{0.39}} & 2.20E-136& \multicolumn{1}{l|}{\textbf{0.35}} & 2.54E-239& \multicolumn{1}{l|}{} &  & \multicolumn{1}{l|}{\textbf{0.28}} & 0.00E+00& \multicolumn{1}{l|}{0.46} & 3.18E-24& \multicolumn{1}{l|}{\textbf{0.28}} & 0.00E+00\\ \hline

\textbf{Random} & \multicolumn{1}{l|}{\textbf{0.62}} & 3.11E-153& \multicolumn{1}{l|}{\textbf{0.64}} & 4.16E-228& \multicolumn{1}{l|}{\textbf{0.62}} & 6.00E-146& \multicolumn{1}{l|}{\textbf{0.72}} & 0.00E+00& \multicolumn{1}{l|}{} &  & \multicolumn{1}{l|}{\textbf{0.69}} & 0.00E+00& \multicolumn{1}{l|}{0.50} & 3.97E-01\\ \hline

\textbf{\DEEPJANUS} & \multicolumn{1}{l|}{\textbf{0.41}} & 1.80E-81& \multicolumn{1}{l|}{\textbf{0.44}} & 2.89E-34& \multicolumn{1}{l|}{\textbf{0.41}} & 2.76E-85& \multicolumn{1}{l|}{0.54} & 3.18E-24& \multicolumn{1}{l|}{\textbf{0.31}} & 0.00E+00& \multicolumn{1}{l|}{} &  & \multicolumn{1}{l|}{\textbf{0.33}} & 0.00E+00\\ \hline

\textbf{\DEEPJANUSGAN} & \multicolumn{1}{l|}{\textbf{0.60}} & 1.45E-116& \multicolumn{1}{l|}{\textbf{0.63}} & 5.03E-178& \multicolumn{1}{l|}{\textbf{0.60}} & 1.90E-103& \multicolumn{1}{l|}{\textbf{0.72}} & 0.00E+00& \multicolumn{1}{l|}{0.50} & 3.97E-01& \multicolumn{1}{l|}{\textbf{0.67}} & 0.00E+00& \multicolumn{1}{l|}{} &  \\ \hline

\end{tabular}%

\vspace{1mm}
Note: tangible (based on $A_{12}$) differences in bold; for each pair of approaches, the best approach is the one with  $A_{12} < 0.50$ and its name on the row. 
\end{table*}

%% file: tables/RQ1_diversity.tex
\begin{table*}[t]
\smaller
\centering
\caption{RQ1-Diversity assessment. Descriptive statistics for diversity across the  generated images for the Urban environment.}
\label{tab:rq1_diversity}
\begin{tabular}{|l|lllllll|}
\hline
\textbf{} & \multicolumn{7}{c|}{\textbf{Feature-based distance}}\\ \hline
\textbf{} & \multicolumn{1}{l|}{\textbf{min}} & \multicolumn{1}{l|}{\textbf{max}} & \multicolumn{1}{l|}{\textbf{median}} & \multicolumn{1}{l|}{\textbf{5th percentile}} & \multicolumn{1}{l|}{\textbf{1st quartile}} & \multicolumn{1}{l|}{\textbf{3rd quartile}} & \textbf{Average} \\ \hline
\textbf{\APPRMONO}  & \multicolumn{1}{l|}{1.15}& \multicolumn{1}{l|}{27.97}  & \multicolumn{1}{l|}{12.83}  & \multicolumn{1}{l|}{9.00}  & \multicolumn{1}{l|}{11.16}  & \multicolumn{1}{l|}{14.71}  & 13.04\\ \hline
\textbf{\APPR} & \multicolumn{1}{l|}{0.48}& \multicolumn{1}{l|}{{28.80}}  & \multicolumn{1}{l|}{\textbf{14.97}}  & \multicolumn{1}{l|}{{9.37}}  & \multicolumn{1}{l|}{\textbf{12.29}}  & \multicolumn{1}{l|}{\textbf{18.08}}  & \textbf{15.25}\\ \hline
\textbf{\APPRPIXEL}& \multicolumn{1}{l|}{0.43}& \multicolumn{1}{l|}{27.58}  & \multicolumn{1}{l|}{12.67}  & \multicolumn{1}{l|}{8.33}  & \multicolumn{1}{l|}{10.90}  & \multicolumn{1}{l|}{14.87}  & 13.05\\ \hline
\textbf{\APPRNOGAN} & \multicolumn{1}{l|}{0.38}& \multicolumn{1}{l|}{15.01}&\multicolumn{1}{l|}{12.90}&  \multicolumn{1}{l|}{5.22}  & \multicolumn{1}{l|}{11.30}& \multicolumn{1}{l|}{13.98} &\multicolumn{1}{l|}{12.08}\\ \hline
\textbf{Random} & \multicolumn{1}{l|}{0.01}& \multicolumn{1}{l|}{12.34}&\multicolumn{1}{l|}{10.99}& \multicolumn{1}{l|}{8.56}& \multicolumn{1}{l|}{10.06}  & \multicolumn{1}{l|}{11.70}& \multicolumn{1}{l|}{10.77}\\ \hline
\textbf{TACTIC} & \multicolumn{1}{l|}{\textbf{8.09}}& \multicolumn{1}{l|}{10.49}  & \multicolumn{1}{l|}{10.11}  & \multicolumn{1}{l|}{9.20}  & \multicolumn{1}{l|}{9.80}& \multicolumn{1}{l|}{10.32}  & 10.06\\ \hline
\textbf{\DEEPJANUS} & \multicolumn{1}{l|}{{2.26}}& \multicolumn{1}{l|}{29.03}& \multicolumn{1}{l|}{13.21}& \multicolumn{1}{l|}{\textbf{10.01}}  & \multicolumn{1}{l|}{11.85}& \multicolumn{1}{l|}{14.78}& 13.45 \\ \hline

\textbf{\DEEPJANUSGAN} & \multicolumn{1}{l|}{{2.66}}& \multicolumn{1}{l|}{\textbf{29.71}}& \multicolumn{1}{l|}{12.11}& \multicolumn{1}{l|}{8.71}  & \multicolumn{1}{l|}{10.60}& \multicolumn{1}{l|}{13.89}& 12.40 \\ \hline

\textbf{} & \multicolumn{7}{c|}{\textbf{Pixel-based distance}}  \\ \hline

\textbf{} & \multicolumn{1}{l|}{\textbf{min}} & \multicolumn{1}{l|}{\textbf{max}} & \multicolumn{1}{l|}{\textbf{median}} & \multicolumn{1}{l|}{\textbf{5th percentile}} & \multicolumn{1}{l|}{\textbf{1st quartile}} & \multicolumn{1}{l|}{\textbf{3rd quartile}} & \textbf{Average} \\ \hline
\textbf{\APPRMONO}  & \multicolumn{1}{l|}{0.30}& \multicolumn{1}{l|}{0.99}& \multicolumn{1}{l|}{0.97}& \multicolumn{1}{l|}{\textbf{0.94}}  & \multicolumn{1}{l|}{0.96}& \multicolumn{1}{l|}{0.98}& \textbf{0.97} \\ \hline
\textbf{\APPR} & \multicolumn{1}{l|}{0.14}& \multicolumn{1}{l|}{\textbf{1.00}}& \multicolumn{1}{l|}{\textbf{0.98}}& \multicolumn{1}{l|}{\textbf{0.94}}  & \multicolumn{1}{l|}{\textbf{0.97}}& \multicolumn{1}{l|}{\textbf{0.99}}& \textbf{0.97} \\ \hline

\textbf{\APPRPIXEL}& \multicolumn{1}{l|}{0.10}& \multicolumn{1}{l|}{\textbf{1.00}}& \multicolumn{1}{l|}{0.97}& \multicolumn{1}{l|}{0.93}  & \multicolumn{1}{l|}{0.96}& \multicolumn{1}{l|}{0.98}& \textbf{0.97} \\ \hline
\textbf{\APPRNOGAN} & \multicolumn{1}{l|}{0.26}&\multicolumn{1}{l|}{0.99}&  \multicolumn{1}{l|}{0.92}  & \multicolumn{1}{l|}{0.83}& \multicolumn{1}{l|}{0.89} &\multicolumn{1}{l|}{0.94}& \multicolumn{1}{l|}{0.91}\\ \hline
\textbf{Random}  & \multicolumn{1}{l|}{0.30}& \multicolumn{1}{l|}{0.76}& \multicolumn{1}{l|}{0.73}& \multicolumn{1}{l|}{0.62}  & \multicolumn{1}{l|}{0.70}& \multicolumn{1}{l|}{0.75}& 0.72 \\ \hline
\textbf{TACTIC} & \multicolumn{1}{l|}{0.43}& \multicolumn{1}{l|}{0.83}& \multicolumn{1}{l|}{0.64}& \multicolumn{1}{l|}{0.51}  & \multicolumn{1}{l|}{0.59}& \multicolumn{1}{l|}{0.68}& 0.63 \\ \hline

\textbf{\DEEPJANUS} & \multicolumn{1}{l|}{\textbf{0.66}}& \multicolumn{1}{l|}{\textbf{1.00}}& \multicolumn{1}{l|}{0.94}& \multicolumn{1}{l|}{0.89}  & \multicolumn{1}{l|}{0.92}& \multicolumn{1}{l|}{0.96}& 0.94 \\ \hline

\textbf{\DEEPJANUSGAN} & \multicolumn{1}{l|}{0.65}& \multicolumn{1}{l|}{\textbf{1.00}}& \multicolumn{1}{l|}{0.97}& \multicolumn{1}{l|}{0.93}  & \multicolumn{1}{l|}{0.95}& \multicolumn{1}{l|}{0.98}& 0.96 \\ \hline

\end{tabular}%

\vspace{1mm}
Note: best (i.e., highest) results (per column) in bold.
\end{table*}

%% file: tables/RQ1_diversity_stats.tex
\begin{table*}[t]
\tiny
\centering
\caption{RQ1-Diversity assessment. p-values and $A_{12}$ for the data in Table~\ref{tab:rq1_diversity}.}
\label{tab:rq1_diversity_stats}
\begin{tabular}{|@{}l|@{}l|@{}l|@{}l|@{}l|@{}l|@{}l|@{}l|@{}l|@{}l|@{}l|@{}l|@{}l|@{}l|@{}l|@{}l|@{}l|@{}l|@{}l|}
\hline
\multicolumn{17}{|c|}{\textbf{Feature-based distance}} \\ \hline
 & \multicolumn{2}{c|}{\APPRMONO} & \multicolumn{2}{c|}{\APPR} & \multicolumn{2}{c|}{\APPRPIXEL} & \multicolumn{2}{c|}{\APPRNOGAN} & \multicolumn{2}{c|}{Random} & \multicolumn{2}{c|}{TACTIC} & \multicolumn{2}{c|}{\DEEPJANUS} & \multicolumn{2}{c|}{\DEEPJANUSGAN} \\ \hline
 
 &  {\VDA{}} &  \textbf{p-value} &  {\VDA{}} &  \textbf{p-value} &  {\VDA{}} &  \textbf{p-value} &  {\VDA{}} &  \textbf{p-value} &  {\VDA{}} &  \textbf{p-value} &  {\VDA{}} &  \textbf{p-value} &  {\VDA{}} &  \textbf{p-value} &  {\VDA{}} &  \textbf{p-value} \\ \hline

\textbf{\APPRMONO} &  {} &  {} &  \textbf{0.33} &  {0.00E+00} &  {0.51} &  {0.00E+00} &  {0.55} &  {0.00E+00} &  \textbf{0.78} &  {0.00E+00} &  \textbf{0.87} &  {0.00E+00} &  {0.45} &  {0.00E+00} &  \textbf{0.57} &  {0.00E+00} \\ \hline

\textbf{\APPR} &  \textbf{0.67} &  {0.00E+00} &  {} &  {} &  \textbf{0.67} &  {0.00E+00} &  \textbf{0.72} &  {0.00E+00} &  \textbf{0.86} &  {0.00E+00} &  \textbf{0.92} &  {0.00E+00} &  \textbf{0.64} &  {0.00E+00} &  \textbf{0.72} &  {0.00E+00} \\ \hline

\textbf{\APPRPIXEL} &  {0.49} &  {0.00E+00} &  \textbf{0.33} &  {0.00E+00} &  {} &  {} &  {0.53} &  {0.00E+00} &  \textbf{0.75} &  {0.00E+00} &  \textbf{0.84} &  {0.00E+00} &  \textbf{0.44} &  {0.00E+00} &  \textbf{0.56} &  {0.00E+00} \\ \hline

\textbf{\APPRNOGAN} &  {0.45} &  {0.00E+00} &  \textbf{0.28} &  {0.00E+00} &  {0.47} &  {0.00E+00} &  {} &  {} &  \textbf{0.77} &  {0.00E+00} &  \textbf{0.84} &  {0.00E+00} &  \textbf{0.40} &  {0.00E+00} &  {0.54} &  {0.00E+00} \\ \hline

\textbf{Random} &  \textbf{0.22} &  {0.00E+00} &  \textbf{0.14} &  {0.00E+00} &  \textbf{0.25} &  {0.00E+00} &  \textbf{0.23} &  {0.00E+00} &  {} &  {} &  \textbf{0.74} &  {5.55E-169} &  \textbf{0.14} &  {0.00E+00} &  \textbf{0.30} &  {0.00E+00} \\ \hline

\textbf{TACTIC} &  \textbf{0.13} &  {0.00E+00} &  \textbf{0.08} &  {0.00E+00} &  \textbf{0.16} &  {0.00E+00} &  \textbf{0.16} &  {0.00E+00} &  \textbf{0.26} &  {5.55E-169} &  {} &  {} &  \textbf{0.06} &  {0.00E+00} &  \textbf{0.18} &  {7.40E-299} \\ \hline

\textbf{\DEEPJANUS} &  {0.55} &  {0.00E+00} &  \textbf{0.36} &  {0.00E+00} &  \textbf{0.56} &  {0.00E+00} &  \textbf{0.60} &  {0.00E+00} &  \textbf{0.86} &  {0.00E+00} &  \textbf{0.94} &  {0.00E+00} &  {} &  {} &  \textbf{0.63} &  {0.00E+00} \\ \hline

\textbf{\DEEPJANUSGAN} &  \textbf{0.43} &  {0.00E+00} &  \textbf{0.28} &  {0.00E+00} &  \textbf{0.44} &  {0.00E+00} &  {0.46} &  {0.00E+00} &  \textbf{0.70} &  {0.00E+00} &  \textbf{0.82} &  {7.40E-299} &  \textbf{0.37} &  {0.00E+00} &  {} &  {} \\ \hline

\multicolumn{17}{|c|}{\textbf{Pixel-based distance}} \\ \hline
 & \multicolumn{2}{c|}{\APPRMONO} & \multicolumn{2}{c|}{\APPR} & \multicolumn{2}{c|}{\APPRPIXEL} & \multicolumn{2}{c|}{\APPRNOGAN} & \multicolumn{2}{c|}{Random} & \multicolumn{2}{c|}{TACTIC} & \multicolumn{2}{c|}{\DEEPJANUS} & \multicolumn{2}{c|}{\DEEPJANUSGAN} \\ \hline

 &  {\VDA{}} &  \textbf{p-value} &  {\VDA{}} &  \textbf{p-value} &  {\VDA{}} &  \textbf{p-value} &  {\VDA{}} &  \textbf{p-value} &  {\VDA{}} &  \textbf{p-value} &  {\VDA{}} &  \textbf{p-value} &  {\VDA{}} &  \textbf{p-value} &  {\VDA{}} &  \textbf{p-value} \\ \hline

\textbf{\APPRMONO} &  {} &  {} &  \textbf{0.39} &  {0.00E+00} &  {0.48} &  {0.00E+00} &  \textbf{0.95} &  {0.00E+00} &  {0.50} &  {1.00E+00} &  \textbf{0.44} &  {3.77E-12} &  \textbf{0.86} &  {0.00E+00} &  \textbf{0.57} &  {0.00E+00} \\ \hline

\textbf{\APPR} &  \textbf{0.61} &  {0.00E+00} &  {} &  {} &  \textbf{0.58} &  {0.00E+00} &  \textbf{0.96} &  {0.00E+00} &  \textbf{0.61} &  {0.00E+00} &  \textbf{0.59} &  {2.74E-22} &  \textbf{0.90} &  {0.00E+00} &  \textbf{0.67} &  {0.00E+00} \\ \hline

\textbf{\APPRPIXEL} &  {0.52} &  {0.00E+00} &  \textbf{0.42} &  {0.00E+00} &  {} &  {} &  {0.94} &  {0.00E+00} &  {0.52} &  {0.00E+00} &  {0.46} &  {3.76E-05} &  \textbf{0.86} &  {0.00E+00} &  \textbf{0.59} &  {0.00E+00} \\ \hline

\textbf{\APPRNOGAN} &  \textbf{0.05} &  {0.00E+00} &  \textbf{0.04} &  {0.00E+00} &  \textbf{0.06} &  {0.00E+00} &  {} &  {} &  \textbf{0.05} &  {0.00E+00} &  \textbf{0.01} &  {0.00E+00} &  \textbf{0.29} &  {0.00E+00} &  \textbf{0.08} &  {0.00E+00} \\ \hline

\textbf{Random} &  {0.50} &  {1.00E+00} &  \textbf{0.39} &  {0.00E+00} &  {0.48} &  {0.00E+00} &  \textbf{0.95} &  {0.00E+00} &  {} &  {} &  \textbf{0.44} &  {3.77E-12} &  \textbf{0.86} &  {0.00E+00} &  \textbf{0.57} &  {0.00E+00} \\ \hline

\textbf{TACTIC} &  \textbf{0.56} &  {3.77E-12} &  \textbf{0.41} &  {2.74E-22} &  \textbf{0.54} &  {3.76E-05} &  \textbf{0.99} &  {0.00E+00} &  \textbf{0.56} &  {3.77E-12} &  {} &  {} &  \textbf{0.96} &  {0.00E+00} &  \textbf{0.65} &  {2.11E-64} \\ \hline

\textbf{\DEEPJANUS} &  \textbf{0.14} &  {0.00E+00} &  \textbf{0.10} &  {0.00E+00} &  \textbf{0.14} &  {0.00E+00} &  \textbf{0.71} &  {0.00E+00} &  \textbf{0.14} &  {0.00E+00} &  \textbf{0.04} &  {0.00E+00} &  {} &  {} &  \textbf{0.19} &  {0.00E+00} \\ \hline

\textbf{\DEEPJANUSGAN} &  \textbf{0.43} &  {0.00E+00} &  \textbf{0.33} &  {0.00E+00} &  \textbf{0.41} &  {0.00E+00} &  \textbf{0.92} &  {0.00E+00} &  \textbf{0.43} &  {0.00E+00} &  \textbf{0.35} &  {2.11E-64} &  \textbf{0.81} &  {0.00E+00} &  {} &  {} \\ \hline

\end{tabular}%

\vspace{1mm}
Note: tangible (based on $A_{12}$) differences in bold; for each pair of approaches, the best approach is the one with  $A_{12} > 0.50$ and its name on the row. 
\end{table*}

%% file: tables/RQ1_diversity_mars.tex
\begin{table*}[t]
\smaller
\centering
\caption{RQ1-Diversity assessment. Descriptive statistics for diversity across the generated images for the Martian environment.}
\label{tab:rq1_diversity_mars}
\begin{tabular}{|l|lllllll|}
\hline

\textbf{} & \multicolumn{7}{c|}{\textbf{Feature Distance Diversity}} \\ \hline
\textbf{} & \multicolumn{1}{l|}{\textbf{min}} & \multicolumn{1}{l|}{\textbf{max}} & \multicolumn{1}{l|}{\textbf{median}} & \multicolumn{1}{l|}{\textbf{5th percentile}} & \multicolumn{1}{l|}{\textbf{1st quartile}} & \multicolumn{1}{l|}{\textbf{3rd quartile}} & \textbf{Average} \\ \hline

\textbf{\APPRMONO} & \multicolumn{1}{l|}{0.00} & \multicolumn{1}{l|}{27.17} & \multicolumn{1}{l|}{10.59} & \multicolumn{1}{l|}{5.67} & \multicolumn{1}{l|}{8.70} & \multicolumn{1}{l|}{12.35} & 10.58 \\ \hline

\textbf{\APPR} & \multicolumn{1}{l|}{0.00} & \multicolumn{1}{l|}{26.24} & \multicolumn{1}{l|}{14.75} & \multicolumn{1}{l|}{8.09} & \multicolumn{1}{l|}{11.22} & \multicolumn{1}{l|}{18.58} & 14.83 \\ \hline

\textbf{\APPRPIXEL} & \multicolumn{1}{l|}{0.82} & \multicolumn{1}{l|}{25.43} & \multicolumn{1}{l|}{11.92} & \multicolumn{1}{l|}{7.79} & \multicolumn{1}{l|}{9.99} & \multicolumn{1}{l|}{14.38} & 12.37 \\ \hline

\textbf{\APPRNOGAN} & \multicolumn{1}{l|}{1.01} & \multicolumn{1}{l|}{\textbf{32.49}} & \multicolumn{1}{l|}{\textbf{16.79}} & \multicolumn{1}{l|}{\textbf{10.23}} & \multicolumn{1}{l|}{\textbf{14.35}} & \multicolumn{1}{l|}{\textbf{19.54}} & \textbf{16.83} \\ \hline

\textbf{Random} & \multicolumn{1}{l|}{0.00} & \multicolumn{1}{l|}{28.70} & \multicolumn{1}{l|}{11.36} & \multicolumn{1}{l|}{7.51} & \multicolumn{1}{l|}{9.44} & \multicolumn{1}{l|}{13.66} & 11.89 \\ \hline

\textbf{\DEEPJANUS} & \multicolumn{1}{l|}{\textbf{2.18}} & \multicolumn{1}{l|}{31.67} & \multicolumn{1}{l|}{12.81} & \multicolumn{1}{l|}{7.16} & \multicolumn{1}{l|}{10.58} & \multicolumn{1}{l|}{18.64} & 14.48 \\ \hline

\textbf{\DEEPJANUSGAN} & \multicolumn{1}{l|}{0.00} & \multicolumn{1}{l|}{25.97} & \multicolumn{1}{l|}{9.52} & \multicolumn{1}{l|}{6.13} & \multicolumn{1}{l|}{7.80} & \multicolumn{1}{l|}{12.25} & 10.54 \\ \hline

\textbf{} & \multicolumn{7}{c|}{\textbf{Pixel Distance Diversity}} \\ \hline
\textbf{} & \multicolumn{1}{l|}{\textbf{min}} & \multicolumn{1}{l|}{\textbf{max}} & \multicolumn{1}{l|}{\textbf{median}} & \multicolumn{1}{l|}{\textbf{5th percentile}} & \multicolumn{1}{l|}{\textbf{1st quartile}} & \multicolumn{1}{l|}{\textbf{3rd quartile}} & \textbf{Average} \\ \hline

\textbf{\APPRMONO} & \multicolumn{1}{l|}{\textbf{0.96}} & \multicolumn{1}{l|}{\textbf{1.00}} & \multicolumn{1}{l|}{0.99} & \multicolumn{1}{l|}{0.97} & \multicolumn{1}{l|}{0.98} & \multicolumn{1}{l|}{0.99} & 0.99 \\ \hline

\textbf{\APPR} & \multicolumn{1}{l|}{0.95} & \multicolumn{1}{l|}{\textbf{1.00}} & \multicolumn{1}{l|}{\textbf{1.00}} & \multicolumn{1}{l|}{\textbf{0.99}} & \multicolumn{1}{l|}{\textbf{1.00}} & \multicolumn{1}{l|}{\textbf{1.00}} & \textbf{1.00} \\ \hline

\textbf{\APPRPIXEL} & \multicolumn{1}{l|}{0.95} & \multicolumn{1}{l|}{\textbf{1.00}} & \multicolumn{1}{l|}{0.98} & \multicolumn{1}{l|}{0.96} & \multicolumn{1}{l|}{0.97} & \multicolumn{1}{l|}{0.99} & 0.98 \\ \hline

\textbf{\APPRNOGAN} & \multicolumn{1}{l|}{0.95} & \multicolumn{1}{l|}{\textbf{1.00}} & \multicolumn{1}{l|}{0.98} & \multicolumn{1}{l|}{0.96} & \multicolumn{1}{l|}{0.97} & \multicolumn{1}{l|}{0.99} & 0.98 \\ \hline

\textbf{Random} & \multicolumn{1}{l|}{\textbf{0.96}} & \multicolumn{1}{l|}{\textbf{\textbf{1.00}}} & \multicolumn{1}{l|}{0.98} & \multicolumn{1}{l|}{0.97} & \multicolumn{1}{l|}{0.98} & \multicolumn{1}{l|}{0.99} & 0.98 \\ \hline

\textbf{\DEEPJANUS} & \multicolumn{1}{l|}{0.95} & \multicolumn{1}{l|}{0.99} & \multicolumn{1}{l|}{0.96} & \multicolumn{1}{l|}{0.95} & \multicolumn{1}{l|}{0.96} & \multicolumn{1}{l|}{0.97} & 0.96 \\ \hline

\textbf{\DEEPJANUSGAN} & \multicolumn{1}{l|}{0.95} & \multicolumn{1}{l|}{\textbf{1.00}} & \multicolumn{1}{l|}{0.99} & \multicolumn{1}{l|}{0.97} & \multicolumn{1}{l|}{0.98} & \multicolumn{1}{l|}{0.99} & 0.98 \\ \hline

\end{tabular}%

\vspace{1mm}
Note: best (i.e., highest or most diverse) results (per column) in bold.
\end{table*}
 

%% file: tables/RQ1_diversity_stats_mars.tex
\begin{table*}[t]
\tiny
\centering
\caption{RQ1-Diversity assessment for the Martian case study. p-values and $A_{12}$ for the data in Table~\ref{tab:rq1_diversity_mars}.}
\label{tab:rq1_diversity_stats_mars}
\begin{tabular}{|@{}l|@{}l@{}p{5mm}@{}l@{}l@{}l@{}l@{}l@{}l@{}l@{}l@{}l@{}l@{}l@{}l@{}l@{}l|}
\hline
\multicolumn{15}{|c|}{\textbf{Feature-based distance}} \\ \hline
 & \multicolumn{2}{l|}{\textbf{\APPRMONO}} & \multicolumn{2}{l|}{\textbf{\APPR}} & \multicolumn{2}{l|}{\textbf{\APPRPIXEL}} & \multicolumn{2}{l|}{\textbf{\APPRNOGAN}} & \multicolumn{2}{l|}{\textbf{Random}} & \multicolumn{2}{l|}{\textbf{\DEEPJANUS}} & \multicolumn{2}{l|}{\textbf{\DEEPJANUSGAN}} \\ \hline

 & \multicolumn{1}{c|}{\textbf{\VDA{}}} & \multicolumn{1}{c|}{\textbf{p-value}} & \multicolumn{1}{c|}{\textbf{\VDA{}}} & \multicolumn{1}{c|}{\textbf{p-value}} & \multicolumn{1}{c|}{\textbf{\VDA{}}} & \multicolumn{1}{c|}{\textbf{p-value}} & \multicolumn{1}{c|}{\textbf{\VDA{}}} & \multicolumn{1}{c|}{\textbf{p-value}} & \multicolumn{1}{c|}{\textbf{\VDA{}}} & \multicolumn{1}{c|}{\textbf{p-value}} & \multicolumn{1}{c|}{\textbf{\VDA{}}} & \multicolumn{1}{c|}{\textbf{p-value}}& \multicolumn{1}{c|}{\textbf{\VDA{}}} & \multicolumn{1}{c|}{\textbf{p-value}} \\ \hline

\textbf{\APPRMONO} & \multicolumn{1}{l|}{} & \multicolumn{1}{l|}{} & \multicolumn{1}{l|}{0.43} & \multicolumn{1}{l|}{0.00E+00} & \multicolumn{1}{l|}{0.52} & \multicolumn{1}{l|}{0.00E+00} & \multicolumn{1}{l|}{0.15} & \multicolumn{1}{l|}{0.00E+00} & \multicolumn{1}{l|}{0.51} & \multicolumn{1}{l|}{0.00E+00} &
\multicolumn{1}{l|}{0.34} & \multicolumn{1}{l|}{0.00E+00} & \multicolumn{1}{l|}{0.62} & \multicolumn{1}{l|}{0.00E+00} \\ \hline

\textbf{\APPR} & \multicolumn{1}{l|}{\textbf{0.57}} & \multicolumn{1}{l|}{0.00E+00} & \multicolumn{1}{l|}{} & \multicolumn{1}{l|}{} & \multicolumn{1}{l|}{\textbf{0.59}} & \multicolumn{1}{l|}{0.00E+00} & \multicolumn{1}{l|}{\textbf{0.19}} & \multicolumn{1}{l|}{0.00E+00} & \multicolumn{1}{l|}{\textbf{0.59}} & \multicolumn{1}{l|}{0.00E+00} & \multicolumn{1}{l|}{\textbf{0.40}} & \multicolumn{1}{l|}{0.00E+00} & \multicolumn{1}{l|}{\textbf{0.68}} & \multicolumn{1}{l|}{0.00E+00} \\ \hline

\textbf{\APPRPIXEL} & \multicolumn{1}{l|}{0.48} & \multicolumn{1}{l|}{0.00E+00} & \multicolumn{1}{l|}{\textbf{0.41}} & \multicolumn{1}{l|}{0.00E+00} & \multicolumn{1}{l|}{} & \multicolumn{1}{l|}{} & \multicolumn{1}{l|}{\textbf{0.13}} & \multicolumn{1}{l|}{0.00E+00} & \multicolumn{1}{l|}{0.50} & \multicolumn{1}{l|}{0.00E+00}& \multicolumn{1}{l|}{\textbf{0.33}} & \multicolumn{1}{l|}{0.00E+00} & \multicolumn{1}{l|}{\textbf{0.62}} & \multicolumn{1}{l|}{0.00E+00} \\ \hline

\textbf{\APPRNOGAN} & \multicolumn{1}{l|}{\textbf{0.85}} & \multicolumn{1}{l|}{0.00E+00} & \multicolumn{1}{l|}{\textbf{0.81}} & \multicolumn{1}{l|}{0.00E+00} & \multicolumn{1}{l|}{\textbf{0.87}} & \multicolumn{1}{l|}{0.00E+00} & \multicolumn{1}{l|}{} & \multicolumn{1}{l|}{} & \multicolumn{1}{l|}{\textbf{0.86}} & \multicolumn{1}{l|}{0.00E+00}& \multicolumn{1}{l|}{\textbf{0.70}} & \multicolumn{1}{l|}{0.00E+00} & \multicolumn{1}{l|}{\textbf{0.88}} & \multicolumn{1}{l|}{0.00E+00} \\ \hline

\textbf{Random} & \multicolumn{1}{l|}{0.49} & \multicolumn{1}{l|}{0.00E+00} & \multicolumn{1}{l|}{\textbf{0.41}} & \multicolumn{1}{l|}{0.00E+00} & \multicolumn{1}{l|}{0.50} & \multicolumn{1}{l|}{0.00E+00} & \multicolumn{1}{l|}{\textbf{0.14}} & \multicolumn{1}{l|}{0.00E+00} & \multicolumn{1}{l|}{} & \multicolumn{1}{l|}{}& \multicolumn{1}{l|}{\textbf{0.33}} & \multicolumn{1}{l|}{0.00E+00} & \multicolumn{1}{l|}{\textbf{0.62}} & \multicolumn{1}{l|}{0.00E+00} \\ \hline

\textbf{\DEEPJANUS} & \multicolumn{1}{l|}{\textbf{0.66}} & \multicolumn{1}{l|}{0.00E+00} & \multicolumn{1}{l|}{\textbf{0.60}} & \multicolumn{1}{l|}{0.00E+00} & \multicolumn{1}{l|}{\textbf{0.67}} & \multicolumn{1}{l|}{0.00E+00} & \multicolumn{1}{l|}{\textbf{0.30}} & \multicolumn{1}{l|}{0.00E+00} & \multicolumn{1}{l|}{\textbf{0.67}} & \multicolumn{1}{l|}{0.00E+00}& \multicolumn{1}{l|}{} & \multicolumn{1}{l|}{} & \multicolumn{1}{l|}{\textbf{0.74}} & \multicolumn{1}{l|}{0.00E+00} \\ \hline

\textbf{\DEEPJANUSGAN} & \multicolumn{1}{l|}{\textbf{0.38}} & \multicolumn{1}{l|}{0.00E+00} & \multicolumn{1}{l|}{\textbf{0.32}} & \multicolumn{1}{l|}{0.00E+00} & \multicolumn{1}{l|}{\textbf{0.38}} & \multicolumn{1}{l|}{0.00E+00} & \multicolumn{1}{l|}{\textbf{0.12}} & \multicolumn{1}{l|}{0.00E+00} & \multicolumn{1}{l|}{\textbf{0.38}} & \multicolumn{1}{l|}{0.00E+00}& \multicolumn{1}{l|}{\textbf{0.26}} & \multicolumn{1}{l|}{0.00E+00} & \multicolumn{1}{l|}{} & \multicolumn{1}{l|}{} \\ \hline

\multicolumn{15}{|c|}{\textbf{Pixel-based distance}} \\ \hline
 & \multicolumn{2}{l|}{\textbf{\APPRMONO}} & \multicolumn{2}{l|}{\textbf{\APPR}} & \multicolumn{2}{l|}{\textbf{\APPRPIXEL}} & \multicolumn{2}{l|}{\textbf{\APPRNOGAN}} & \multicolumn{2}{l|}{\textbf{Random}} & \multicolumn{2}{l|}{\textbf{\DEEPJANUS}} & \multicolumn{2}{l|}{\textbf{\DEEPJANUSGAN}} \\ \hline

& \multicolumn{1}{c|}{\textbf{\VDA{}}} & \multicolumn{1}{c|}{\textbf{p-value}} & \multicolumn{1}{c|}{\textbf{\VDA{}}} & \multicolumn{1}{c|}{\textbf{p-value}} & \multicolumn{1}{c|}{\textbf{\VDA{}}} & \multicolumn{1}{c|}{\textbf{p-value}} & \multicolumn{1}{c|}{\textbf{\VDA{}}} & \multicolumn{1}{c|}{\textbf{p-value}} & \multicolumn{1}{c|}{\textbf{\VDA{}}} & \multicolumn{1}{c|}{\textbf{p-value}} & \multicolumn{1}{c|}{\textbf{\VDA{}}} & \multicolumn{1}{c|}{\textbf{p-value}}& \multicolumn{1}{c|}{\textbf{\VDA{}}} & \multicolumn{1}{c|}{\textbf{p-value}} \\ \hline

\textbf{\APPRMONO} & \multicolumn{1}{l|}{} & \multicolumn{1}{l|}{} & \multicolumn{1}{l|}{\textbf{0.36}} & \multicolumn{1}{l|}{2.23E-211} & \multicolumn{1}{l|}{\textbf{0.60}} & \multicolumn{1}{l|}{3.56E-106} & \multicolumn{1}{l|}{\textbf{0.81}} & \multicolumn{1}{l|}{0.00E+00} & \multicolumn{1}{l|}{0.55} & \multicolumn{1}{l|}{1.08E-26} & \multicolumn{1}{l|}{\textbf{0.99}} & \multicolumn{1}{l|}{0.00E+00} & \multicolumn{1}{l|}{0.55} & \multicolumn{1}{l|}{1.32E-35}\\ \hline

\textbf{\APPR} & \multicolumn{1}{l|}{\textbf{0.64}} & \multicolumn{1}{l|}{2.23E-211} & \multicolumn{1}{l|}{} & \multicolumn{1}{l|}{} & \multicolumn{1}{l|}{\textbf{0.71}} & \multicolumn{1}{l|}{0.00E+00} & \multicolumn{1}{l|}{\textbf{0.83}} & \multicolumn{1}{l|}{0.00E+00} & \multicolumn{1}{l|}{\textbf{0.68}} & \multicolumn{1}{l|}{0.00E+00} & \multicolumn{1}{l|}{\textbf{0.93}} & \multicolumn{1}{l|}{0.00E+00} & \multicolumn{1}{l|}{\textbf{0.68}} & \multicolumn{1}{l|}{0.00E+00}\\ \hline

\textbf{\APPRPIXEL} & \multicolumn{1}{l|}{\textbf{0.40}} & \multicolumn{1}{l|}{3.56E-106} & \multicolumn{1}{l|}{\textbf{0.29}} & \multicolumn{1}{l|}{0.00E+00} & \multicolumn{1}{l|}{} & \multicolumn{1}{l|}{} & \multicolumn{1}{l|}{\textbf{0.77}} & \multicolumn{1}{l|}{0.00E+00} & \multicolumn{1}{l|}{\textbf{0.44}} & \multicolumn{1}{l|}{4.51E-40}& \multicolumn{1}{l|}{\textbf{0.98}} & \multicolumn{1}{l|}{0.00E+00} & \multicolumn{1}{l|}{0.45} & \multicolumn{1}{l|}{0.00E+00} \\ \hline

\textbf{\APPRNOGAN} & \multicolumn{1}{l|}{\textbf{0.19}} & \multicolumn{1}{l|}{0.00E+00} & \multicolumn{1}{l|}{\textbf{0.17}} & \multicolumn{1}{l|}{0.00E+00} & \multicolumn{1}{l|}{\textbf{0.23}} & \multicolumn{1}{l|}{0.00E+00} & \multicolumn{1}{l|}{} & \multicolumn{1}{l|}{} & \multicolumn{1}{l|}{\textbf{0.20}} & \multicolumn{1}{l|}{0.00E+00} & \multicolumn{1}{l|}{\textbf{0.88}} & \multicolumn{1}{l|}{0.00E+00} & \multicolumn{1}{l|}{\textbf{0.21}} & \multicolumn{1}{l|}{0.00E+00}\\ \hline

\textbf{Random} & \multicolumn{1}{l|}{0.45} & \multicolumn{1}{l|}{1.08E-26} & \multicolumn{1}{l|}{\textbf{0.32}} & \multicolumn{1}{l|}{0.00E+00} & \multicolumn{1}{l|}{\textbf{0.56}} & \multicolumn{1}{l|}{4.51E-40} & \multicolumn{1}{l|}{\textbf{0.80}} & \multicolumn{1}{l|}{0.00E+00} & \multicolumn{1}{l|}{} & \multicolumn{1}{l|}{}& \multicolumn{1}{l|}{\textbf{0.99}} & \multicolumn{1}{l|}{0.00E+00} & \multicolumn{1}{l|}{0.51} & \multicolumn{1}{l|}{0.00E+00} \\ \hline

\textbf{\DEEPJANUS} & \multicolumn{1}{l|}{\textbf{0.01}} & \multicolumn{1}{l|}{0.00E+00} & \multicolumn{1}{l|}{\textbf{0.07}} & \multicolumn{1}{l|}{0.00E+00} & \multicolumn{1}{l|}{\textbf{0.02}} & \multicolumn{1}{l|}{0.00E+00} & \multicolumn{1}{l|}{\textbf{0.12}} & \multicolumn{1}{l|}{0.00E+00} & \multicolumn{1}{l|}{\textbf{0.01}} & \multicolumn{1}{l|}{0.00E+00} & \multicolumn{1}{l|}{} & \multicolumn{1}{l|}{} & \multicolumn{1}{l|}{\textbf{0.01}} & \multicolumn{1}{l|}{0.00E+00}\\ \hline

\textbf{\DEEPJANUSGAN} & \multicolumn{1}{l|}{0.45} & \multicolumn{1}{l|}{1.32E-35} & \multicolumn{1}{l|}{\textbf{0.32}} & \multicolumn{1}{l|}{0.00E+00} & \multicolumn{1}{l|}{0.55} & \multicolumn{1}{l|}{1.11E-34} & \multicolumn{1}{l|}{\textbf{0.79}} & \multicolumn{1}{l|}{0.00E+00} & \multicolumn{1}{l|}{0.49} & \multicolumn{1}{l|}{1.19E-01}& \multicolumn{1}{l|}{\textbf{0.99}} & \multicolumn{1}{l|}{0.00E+00} & \multicolumn{1}{l|}{} & \multicolumn{1}{l|}{} \\ \hline

\end{tabular}
\vspace{1mm}
\\
Note: tangible (based on $A_{12}$) differences in bold; for each pair of approaches, the best approach is the one with  $A_{12} < 0.50$ and its name on the row. 
\end{table*}

%% file: tables/RQ2_results.tex
\begin{table*}[t]
\centering
\footnotesize
\caption{RQ2. $IoU_{car}$ results obtained with the DeeplabV3 model retrained using the outputs of the different approaches.}
\label{tab:rq2_results}
\begin{tabular}{|l|l|l|l|l|l|l|l|}
\hline
 & \multicolumn{7}{c|}{\textbf{$\mathbf{IoU_{car}}$ distribution}} \\ \hline
\textbf{Retraining set}& \textbf{min}  & \textbf{max}  & \textbf{median} & \textbf{5th percentile} & \textbf{1st quartile} & \textbf{3rd quartile} & \textbf{average} \\ \hline
\textbf{None (original DeeplabV3)} & 0.00 & 0.97 & 0.82 & 0.00 & 0.62 & 0.91 & 0.71 \\ \hline
\textbf{Cityscapes (Retrained DNN)} & 0.00 (+0.00) & 0.97 (+0.00) & 0.84 (+0.02) & 0.00 (+0.00) & 0.70 (+0.08) & 0.91 (+0.00) & 0.74 (+0.03) \\ \hline
\textbf{\APPRMONO} & 0.00 (+0.00) & \textbf{0.98 (+0.01)} & 0.86 (+0.04) & 0.00 (+0.00) & 0.73 (+0.11) & 0.92 (+0.01) & 0.76 (+0.05) \\ \hline
\textbf{\APPR} & 0.00 (+0.00) & \textbf{0.98 (+0.01)} & \textbf{0.90 (+0.08)} & \textbf{0.34 (+0.34)} & \textbf{0.80 (+0.18)} & \textbf{0.94 (+0.03)} & \textbf{0.82 (+0.11)} \\ \hline
\textbf{\APPRPIXEL} & 0.00 (+0.00) & \textbf{0.98 (+0.01)} & 0.85 (+0.03) & 0.00 (+0.00) & 0.70 (+0.08) & 0.92 (+0.01) & 0.75 (+0.04) \\ \hline
\textbf{\APPRNOGAN} & 0.00 (+0.00) & 0.97 (+0.00) & 0.86 (+0.04) & 0.00 (+0.00) & 0.73 (+0.1) & 0.92 (+0.01) & 0.76 (+0.05) \\ \hline
\textbf{Random} & 0.00 (+0.00) & \textbf{0.98 (+0.01)} & 0.85 (+0.03) & 0.00 (+0.00) & 0.70 (+0.08) & 0.92 (+0.01) & 0.74 (+0.03) \\ \hline
\textbf{TACTIC} & 0.00 (+0.00) & 0.97 (+0.00) & 0.84 (+0.02) & 0.00 (+0.00) & 0.69 (+0.07) & 0.92 (+0.01) & 0.74 (+0.03) \\ \hline
\textbf{\DEEPJANUS}& 0.00 (+0.00)& \textbf{0.98 (+0.01)}& 0.87 (+0.05)& 0.00 (+0.00)& 0.75 (+0.13)& 0.93 (+0.02)& 0.77 (+0.06)\\ \hline
\textbf{\DEEPJANUSGAN}& 0.00 (+0.00)& 0.97 (+0.00)& 0.86 (+0.04)& 0.00 (+0.00)& 0.73 (+0.11)& 0.92 (+0.01)& 0.76 (+0.05)\\ \hline
\end{tabular}%
\vspace{1mm}
\\
Note: best (i.e., highest or most diverse) results (per column) in bold.
\end{table*}

%% file: tables/RQ2_stats.tex
\begin{table*}[t]
\centering
\caption{RQ2. p-values and $A_{12}$ for the data in Table~\ref{tab:rq2_results}.}
\label{tab:rq2_stats}
\resizebox{\textwidth}{!}{%
\begin{tabular}{|@{}l|@{}l|@{}l|@{}l|@{}l|@{}l|@{}l|@{}l|@{}l|@{}l|@{}l|@{}l|@{}l|@{}l|@{}l|@{}l|@{}l|@{}l|@{}l|l|l|}
\hline
 {}                          & \multicolumn{2}{|c|}{\textbf{Pre-trained   DeeplabV3}} & \multicolumn{2}{|c|}{\textbf{Retrained DeeplabV3}} & \multicolumn{2}{|c|}{\textbf{\APPRMONO}} & \multicolumn{2}{|c|}{\textbf{\APPR}} & \multicolumn{2}{|c|}{\textbf{\APPRPIXEL}} & \multicolumn{2}{|c|}{\textbf{\APPRNOGAN}} & \multicolumn{2}{|c|}{\textbf{Random}} & \multicolumn{2}{|c|}{\textbf{TACTIC}} &   \multicolumn{2}{|c|}{\textbf{\DEEPJANUS}} & \multicolumn{2}{|c|}{\textbf{\DEEPJANUSGAN}}\\ \hline
\multicolumn{1}{|l|}{}                          & \multicolumn{1}{|l|}{\VDA{}}      & p-value      & \multicolumn{1}{|l|}{\VDA{}}     & p-value   & \multicolumn{1}{|l|}{\VDA{}}       & p-value      & \multicolumn{1}{|l|}{\VDA{}}     & p-value    & \multicolumn{1}{|l|}{\VDA{}}       & p-value       & \multicolumn{1}{|l|}{\VDA{}}            &      p-value        & \multicolumn{1}{|l|}{\VDA{}}   & p-value   & VDA                         & p-value                & VDA                         & p-value                & VDA                         &p-value                \\ \hline

{\textbf{Pre-trained DeeplabV3}}                                     &&              &  {0.47}    & 0.06      &  {\textbf{0.44}}      & 0.00E+00     &  {\textbf{0.35}}    & 0.00E+00   &  {0.46}      & 0.00E+00      &  {\textbf{0.44}}            &    0.00E+00         &  {0.46}   &  0.00E+00 & 0.47                        & 2.00E-02               & 0.42& 0.00E+00& \textbf{0.44}&0.00E+00\\ \hline

 {\textbf{Retrained DeeplabV3}}       &  {0.53}     & 6.00E-02     &  {}        &           &  {0.47}      & 0.00E+00     &  \textbf{0.37}    & 0.00E+00   &  {0.49}      & 5.00E-02      &  \textbf{0.46}            &      0.00E+00       &  {0.49}   &  6,00E-02 & 0.49                        & 5.00E-01               & \textbf{0.44}& 0.00E+00& 0.47&0.00E+00\\ \hline

 {\textbf{\APPRMONO}}  &  {\textbf{0.56}}     & 0.00E+00     &  {0.53}    & 0.00      &  {}          &              &  {\textbf{0.40}}    & 0.00E+00   &  {0.52}      & 0.00E+00      &  {0.49}            &    0.00E+00         &  {0.52}   & 0,00E+00  & 0.53                        & 0.00E+00               & 0.47& 0.00E+00& 0.50&7.00E-01\\ \hline

 {\textbf{\APPR}}      &  {\textbf{0.65}}     & 0.00E+00     &  {\textbf{0.63}}    & 0.00      &  {\textbf{0.60}}      & 0.00E+00     &  {}        &            &  {\textbf{0.61}}      & 0.00E+00      &  {\textbf{0.59}}            &     0.00E+00        &  {\textbf{0.61}}   & 0,00E+00  & \textbf{0.62}                        & 0.00E+00               & \textbf{0.57}& 0.00E+00& \textbf{0.59}&0.00E+00\\ \hline

 {\textbf{\APPRPIXEL}} &  {0.54}     & 0.00E+00     &  {0.51}    & 0.05      &  {0.48}      & 0.00E+00     &  {\textbf{0.39}}    & 0.00E+00   &  {}          &               &  {0.48}            &      0.00E+00       &  {0.50}   & 8,40E-01  & 0.51                        & 1.40E-01               & 0.45& 0.00E+00& 0.48&0.00E+00\\ \hline

 {\textbf{\APPRNOGAN}} &  {\textbf{0.56}}         &       0.00E+00       &  {0.54}        &     0.00E+00      &  {0.51}         &        1.00E-01       &  \textbf{0.41}        &    0.00E+00        &  {0.52}          &       0.00E+00        &  {}            &             &  {0.53}   & 0.00E+00  &         0.53                    &        0.00E+00                & 0.48& 0.00E+00& 0.51&2.70E-01\\ \hline

 {\textbf{Random}}                    &  {0.54}                          & 0.00E+00                         &  {0.51}                        & 6.00E-02                       &  {0.48}                          & 0.00E+00                          &  {\textbf{0.39}}                        & 0,00E+00                        &  {0.50}                            & 8,40E-01                          &  {0.47}                               &                0.00E+00                   &  {}                      &                         & 0.51                                            & 1.90E-01  & 0.45& 0.00E+00& 0.48&0.00E+00\\ \hline

 {\textbf{TACTIC}}                    &  {0.53}     & 2.00E-02     &  {0.51}    & 0.50      &  {0.47}      & 0.00E+00     &  \textbf{0.38}    & 0.00E+00   &  {0.49}      & 1.40E-01      &  {0.47}            &      0.00E+00       &  {0.49}   &  1,90E-01 &                             &                        & \textbf{0.44}& 0.00E+00& 0.47&0.00E+00\\ \hline
 \textbf{\DEEPJANUS}& 
\textbf{0.58}& 0.00E+00& \textbf{0.56}& 0.00E+00& 0.53& 0.00E+00& \textbf{0.43}& 0.00E+00& 0.55& 0.00E+00& 0.52& 0.00E+00& 0.55& 0.00E+00& \textbf{0.56}& 0.00E+00& & & 0.53&0.00E+00\\ \hline 
 \textbf{\DEEPJANUSGAN}& \textbf{0.56}& 0.00E+00& 0.53& 0.00E+00& 0.50& 7.00E-01& \textbf{0.41}& 0.00E+00& 0.52& 0.00E+00& 0.49& 2.70E-01& 0.52& 0.00E+00& 0.53& 0.00E+00& 0.47& 0.00E+00& &\\ \hline
\end{tabular}%
}
\vspace{1mm}
\\
Note: tangible (based on $A_{12}$) differences in bold; for each pair of approaches, the best approach is the one with  $A_{12} > 0.50$ and its name on the row. 
\end{table*}

%% file: tables/RQ2_results_mars.tex
\begin{table*}[t]
\centering
\footnotesize
\caption{RQ2. $mIoU$ results obtained with the DeeplabV3 model retrained using the outputs of the different approaches for the Martian environment.}
\label{tab:rq2_results_mars}
\begin{tabular}{|@{}l|l|l|l|l|l|l|l|}
\hline
\textbf{Retraining set} & \multicolumn{1}{l|}{\textbf{min}} & \multicolumn{1}{l|}{\textbf{max}} & \multicolumn{1}{l|}{\textbf{median}} & \multicolumn{1}{l|}{\textbf{5th percentile}} & \multicolumn{1}{l|}{\textbf{1st quartile}} & \multicolumn{1}{l|}{\textbf{3rd quartile}} & \textbf{Average} \\ \hline

\textbf{None (Original DeeplabV3)} & 0.00 & 0.99 & 0.44 & 0.01 & 0.26 & \textbf{0.77} & 0.49 \\ \hline
\textbf{AI4Mars training (Retrained DNN)} & \textbf{0.01 (+0.01) }& 1.00 (+0.01) & 0.47 (+0.03) & 0.17 (+0.16) & 0.33 (+0.07) & 0.63 (-0.14) & 0.49 (+0.00) \\ \hline
\textbf{\APPRMONO} & 0.00 (+0.00) & 1.00 (+0.01) & 0.50 (+0.06) & 0.19 (+0.18) & 0.34 (+0.08) & 0.67 (-0.10) & 0.52 (+0.03) \\ \hline
\textbf{\APPR} & 0.00 (+0.00) & 1.00 (+0.01) & \textbf{0.53 (+0.09)} & 0.19 (+0.18) & \textbf{0.38 (+0.12)} & 0.69 (-0.08) & \textbf{0.53 (+0.04)} \\ \hline
\textbf{\APPRPIXEL} & 0.00 (+0.00) & 1.00 (+0.01) & 0.50 (+0.06) & 0.15 (+0.14) & 0.36 (+0.09) & 0.68 (-0.09) & 0.52 (+0.02) \\ \hline
\textbf{\APPRNOGAN} & 0.00 (+0.00) & 1.00 (+0.01) & 0.46 (+0.02) & 0.07 (+0.06) & 0.30 (+0.04) & 0.64 (-0.13) & 0.47 (-0.02) \\ \hline
\textbf{Random} & 0.00 (+0.00) & 1.00 (+0.01) & 0.49 (+0.05) & 0.12 (+0.10) & 0.33 (+0.07) & 0.68 (-0.09) & 0.51 (+0.02) \\ \hline
\textbf{\DEEPJANUS} & 0.00 (+0.00) & 1.00 (+0.01) & 0.50 (+0.06) & 0.19 (+0.18) & 0.36 (+0.09) & 0.66 (-0.10) & 0.52 (+0.03) \\ \hline
\textbf{\DEEPJANUSGAN} & 0.00 (+0.00) & 1.00 (+0.01) & 0.50 (+0.06) & \textbf{0.20 (+0.19)} & 0.36 (+0.09) & 0.67 (-0.10) & 0.52 (+0.03) \\ \hline
\end{tabular}%
\vspace{1mm}
\\
Note: best (i.e., highest or most diverse) results (per column) in bold.
\end{table*} 

%% file: tables/RQ2_stats_mars.tex
\begin{table*}[t]
\centering
\tiny
\caption{RQ2. p-values and $A_{12}$ for the data in Table~\ref{tab:rq2_results_mars}}.
\label{tab:rq2_stats_mars}
\resizebox{\textwidth}{!}{%
\begin{tabular}{|l|l|l|l|l|l|l|l|l|l|l|l|l|l|l|l|l|l|l|}
\hline
\multicolumn{1}{|l|}{}                          & \multicolumn{2}{|c|}{\textbf{Original}} & \multicolumn{2}{|c|}{\textbf{Original\_retrained}} & \multicolumn{2}{|c|}{\textbf{\APPRMONO}} & \multicolumn{2}{|c|}{\textbf{\APPR}} & \multicolumn{2}{|c|}{\textbf{\APPRPIXEL}} & \multicolumn{2}{|c|}{\textbf{\APPRNOGAN}} & \multicolumn{2}{|c|}{\textbf{Random}} &  \multicolumn{2}{|c|}{\textbf{\DEEPJANUS}} & \multicolumn{2}{|c|}{\textbf{\DEEPJANUSGAN}}\\ \hline
\multicolumn{1}{|l|}{}                          & \multicolumn{1}{|l|}{\textbf{\VDA{}}}      & \textbf{p-value}      & \multicolumn{1}{|l|}{\textbf{\VDA{}}}     & \textbf{p-value}   & \multicolumn{1}{|l|}{\textbf{\VDA{}}}       & \textbf{p-value}      & \multicolumn{1}{|l|}{\textbf{\VDA{}}}     & \textbf{p-value}    & \multicolumn{1}{|l|}{\textbf{\VDA{}}}       & \textbf{p-value}       & \multicolumn{1}{|l|}{\textbf{\VDA{}}}            & \textbf{p-value}        & \multicolumn{1}{|l|}{\textbf{\VDA{}}}   & \textbf{p-value}    & \textbf{\VDA{}}&\textbf{p-value}     & &\\ \hline

\multicolumn{1}{|l|}{\textbf{Original}}  & \multicolumn{1}{|l|}{}     &   & \multicolumn{1}{|l|}{0.47}    & 1.60E-01& \multicolumn{1}{|l|}{0.47}      & 1.00E-02& \multicolumn{1}{|l|}{0.39}    & 0.00E+00& \multicolumn{1}{|l|}{0.43}      & 0.00E+00& \multicolumn{1}{|l|}{0.51}            &    5.70E-01& \multicolumn{1}{|l|}{0.48}   &  1.60E-01& \textbf{0.44}& 0.00E+00& \textbf{0.44}&0.00E+00\\ \hline

\multicolumn{1}{|l|}{\textbf{Retrained DNN}}       & \multicolumn{1}{|l|}{0.53}     & 1.60E-01& \multicolumn{1}{|l|}{}        &           & \multicolumn{1}{|l|}{0.50}      & 7.40E-01& \multicolumn{1}{|l|}{0.44}    & 0.00E+00& \multicolumn{1}{|l|}{0.48}      & 2.30E-01& \multicolumn{1}{|l|}{0.54}            & 0.00E+00& \multicolumn{1}{|l|}{0.52}   &  1.80E-01& 0.49& 3.90E-01& 0.49&3.10E-01\\ \hline

\multicolumn{1}{|l|}{\textbf{\APPRMONO}}       & \multicolumn{1}{|l|}{0.53}     & 1.00E-02& \multicolumn{1}{|l|}{0.50}    & 7.40E-01& \multicolumn{1}{|l|}{}      & & \multicolumn{1}{|l|}{0.43}    & 0.00E+00& \multicolumn{1}{|l|}{0.47}      & 0.00E+00& \multicolumn{1}{|l|}{0.54}            & 0.00E+00& \multicolumn{1}{|l|}{0.51}   &  3.00E-02& 0.48& 0.00E+00& 0.48&0.00E+00\\ \hline

\multicolumn{1}{|l|}{\textbf{\APPR}}       & \multicolumn{1}{|l|}{\textbf{0.61}}     & 0.00E+00& \multicolumn{1}{|l|}{\textbf{0.56}}    & 0.00E+00& \multicolumn{1}{|l|}{\textbf{0.57}}      & 0.00E+00& \multicolumn{1}{|l|}{}    & & \multicolumn{1}{|l|}{0.54}      & 0.00E+00& \multicolumn{1}{|l|}{\textbf{0.61}}            & 0.00E+00& \multicolumn{1}{|l|}{\textbf{0.58}}   &  0.00E+00& 0.55& 0.00E+00& 0.55&0.00E+00\\ \hline

\multicolumn{1}{|l|}{\textbf{\APPRPIXEL}}       & \multicolumn{1}{|l|}{\textbf{0.57}}     & 0.00E+00& \multicolumn{1}{|l|}{0.52}    & 2.30E-01& \multicolumn{1}{|l|}{0.53}      & 0.00E+00& \multicolumn{1}{|l|}{0.46}    & 0.00E+00& \multicolumn{1}{|l|}{}      & & \multicolumn{1}{|l|}{\textbf{0.57}}            & 0.00E+00& \multicolumn{1}{|l|}{0.54}   &  0.00E+00& 0.51& 1.00E-02& 0.51&5.00E-02\\ \hline
\multicolumn{1}{|l|}{\textbf{\APPRNOGAN}}       & \multicolumn{1}{|l|}{0.49}     & 5.70E-01& \multicolumn{1}{|l|}{0.46}    & 0.00E+00& \multicolumn{1}{|l|}{0.46}      & 0.00E+00& \multicolumn{1}{|l|}{\textbf{0.39}}    & 0.00E+00& \multicolumn{1}{|l|}{\textbf{0.43}}      & 0.00E+00& \multicolumn{1}{|l|}{}            &       & \multicolumn{1}{|l|}{0.47}   &  0.00E+00& \textbf{0.44}& 0.00E+00& \textbf{0.44}&0.00E+00\\ \hline
\multicolumn{1}{|l|}{\textbf{Random}}       & \multicolumn{1}{|l|}{0.52}     & 1.60E-01& \multicolumn{1}{|l|}{0.48}    & 1.80E-01& \multicolumn{1}{|l|}{0.49}      & 3.00E-02& \multicolumn{1}{|l|}{\textbf{0.42}}    & 0.00E+00& \multicolumn{1}{|l|}{0.46}      & 0.00E+00& \multicolumn{1}{|l|}{0.53}            & 0.00E+00& \multicolumn{1}{|l|}{}   &   & 0.47& 0.00E+00& 0.47&0.00E+00\\ \hline
 \textbf{\DEEPJANUS}& \textbf{0.56}& 0.00E+00& 0.51& 3.90E-01& 0.52& 0.00E+00& 0.45& 0.00E+00& 0.49& 1.00E-02& \textbf{0.56}& 0.00E+00& 0.53& 0.00E+00& & & 0.50&6.30E-01\\\hline
 \textbf{\DEEPJANUSGAN}& \textbf{0.56}& 0.00E+00& 0.51& 3.10E-01& 0.52& 0.00E+00& 0.45& 0.00E+00& 0.49& 5.00E-02& \textbf{0.56}& 0.00E+00& 0.53& 0.00E+00& 0.50& 6.30E-01& &\\\hline
\end{tabular}%
}
\vspace{1mm}\\
Note: tangible (based on $A_{12}$) differences in bold; for each pair of approaches, the best approach is the one with  $A_{12} > 0.50$ and its name on the row. 
\end{table*}

%% file: related.tex
\section{Related work}
\label{sec:related}

Our work relates to DNN testing approaches leveraging generative neural networks, which are mainly adopted to perform adversarial testing or generating test images by altering image elements.

Adversarial testing technique have a different purpose than \APPR; indeed, they aim to assess whether DNNs provide correct predictions when attackers purposely alter inputs. \APPR, instead, aims to ensure correct DNN behaviour within its operational design domain~\cite{Zhang2023}. The state of the art technique is FuzzGan~\cite{han2022fuzzgan}, which relies on an auxiliary classifier GAN~\cite{ACGAN} that is trained to learn the representation of images belonging to specific classes and triggering specific neuron activations. During testing, the GAN is used to generate inputs that likely triggers certain neuron activations; testing terminates when the desired neuron coverage is achieved~\cite{Canada}. FuzzGan outperforms
CAGFuzz \cite{zhang2021cagfuzz}, an approach relying on CycleGAN for the same purpose. An alternative approach to adversarial testing has been recently proposed by Yuan et al.; it consists of performing search-based exploration on a manifold obtained through a GAN~\cite{ManifoldExploration}. Yuan's approach generates valid images, as opposed to approaches mutating the latent space captured by variational autoencoders~\cite{SINVAD:Yoo:Feldt,Dola2021}; further, it outperforms other DNN testing approaches (DeepHunter~\cite{xie2019deephunter}, TensorFuzz~\cite{pmlr-v97-odena19a}, DeepTest~\cite{Tian2017DeepTest}) in terms of number of reported failures, and it is the only approach not reducing accuracy after retraining but slightly increasing it. Such results show that inputs generated by adversarial attacks tend to reduce accuracy, thus not being an appropriate solution to our problem. \APPR, in contrast to Yuan's approach, does not focus on adversarial testing, but aims at improving  DNN accuracy. Finally, \APPR can be applied to classifier and regression DNNs, while Yuan's approach can only be applied to the former because, for regression tasks, it is not possible to automatically derive a ground truth for the generated images. 

Inspired by metamorphic testing approaches relying on image modifications~\cite{Tian2017DeepTest,pei2017deepxplore}, several techniques rely on GANs to change the weather conditions in input images, and report failures when the DNN output differs from the original inputs. For example, AdversarialStyle relies on MUNIT~\cite{AdversarialStyle}, while DeepRoad relies on UNIT~\cite{zhang2018deeproad}. The state-of-the-art approach is TACTIC~\cite{li2021testing}, which relies on MUNIT to introduce environmental changes (see Section~\ref{sec:empirical:techniques}) and, in addition, relies on a multi-objective search approach to further alter the MUNIT style vector to generate images that are diverse (using a fitness based on neuron coverage) and leading to the worst prediction error.
Unfortunately, TACTIC can alter only the environmental conditions of existing images, which limits input diversity, while \APPR can leverage a simulator to generate inputs that cannot be obtained by altering a given image (e.g., rotating the view). Such difference between \APPR and TACTIC leads to a much better accuracy obtained by \APPR during retraining, as demonstrated by our empirical assessment.

Without leveraging GANs, \emph{semInFuzz}~\cite{woodlief2022semantic} relies on the segmentation masks provided by Cityscapes to identify objects (e.g., a car) to be extracted from seed images and paste them into the images to be used as test inputs. Being based on simple transformations (e.g., cut and paste), it may lead to unrealistic images (e.g., proportions, perspective, or object blending). 


A large number of approaches rely on meta-heuristic search to drive simulators for cost effective testing~\cite{fitash:offline:emse,Haq:2021,abdessalem2018testing,Riccio2021,Hazem:SEDE,Gambi2019}, with some recent solutions maximizing input diversity to maximize the number of diverse fault being detected~\cite{riccio2020model,zohdinasab2021deephyperion,Zohdinasab2023,Zohdinasab2023b}. Recently, reinforcement learning has also shown to be an effective solution for DNN testing~\cite{MORLOT}. However, all these approaches do not lead to realistic images because they do not rely on GANs; instead, we demonstrated that the combination of GANs with meta-heuristic search, leveraging simulators, leads to better results for both testing and retraining of vision DNNs.

To the best of our knowledge, \APPR is the first technique combining meta-heuristic search, simulation, and GANs, to test and retrain DNNs for safety-critical tasks, based on realistic images.
In other contexts, preliminary works report on the feasibility of relying on CycleGAN to generate, from simulator outputs, realistic sea images~\cite{DONG2023103456}, sonar data~\cite{Liu}, and landscapes~\cite{Arellano2024}; other work proposes testing object detection DNNs by applying GANs to maritime simulator outputs but do empirically assess the idea~\cite{Osen}. In the software engineering context, the closer work is that of Stocco et al. \cite{stocco2022mind}, who evaluated the 
differences in testing results obtained in a simulated environment versus those from a physical setting, with a small-size vehicle (i.e., Donkeycar~\cite{Donkeycar}). They use CycleGAN to generate realistic images from simulated driving scenes and train a telemetry predictor. In line with our findings, their results show that images generated by GANs can be used to successfully train a DNN used in the physical world. However, because of the nature of their study, they do not demonstrate that simulators and GANs can be integrated within a search-based process; instead, our work demonstrates how to perform such integration in an effective way, without leading to the generation of images that fail simply because unrealistic. 

%% file: conclusion.tex
\section{Conclusion}
\label{sec:conclusion}
In this paper, we proposed \APPR, a novel approach that integrates GANs with meta-heuristic search and simulators to enhance the testing and retraining of DNNs. By leveraging GANs, \APPR transforms simulator-generated images into high-fidelity, realistic inputs that closely resemble real-world data distributions. This transformation addresses the fidelity gap inherent in simulator-based testing, ensuring that the observed DNN failures are indicative of real-world performance limitations.

Our empirical investigation, conducted using a state-of-the-art DNN for road and martian terrains segmentation, demonstrates the effectiveness of \APPR in identifying failure-inducing inputs and maximizing input diversity. \APPR consistently outperformed the baseline approaches, including the state-of-the-art methods like TACTIC and \DEEPJANUS, by generating inputs that led to worse DNN performance and more diverse failure scenarios. Furthermore, the realistic images generated by \APPR proved to be valuable for DNN retraining, leading to improved performance on real-world images, thus outperforming alternative solutions and demonstrating that: (1) simulator images are not as effective as realistic images for DNN retraining (2) relying on GANs to generate images that simply change weather conditions is not as effective as combining GAN with a simulator to generate realistic images for DNN retraining. Based on our findings, we suggest that future work on DNN testing, instead of simply relying on simulators, should integrate GAN components transforming simulator outputs into more realistic ones.

%% file: main.bbl
\begin{thebibliography}{10}

\bibitem{abdessalem2018testing}
Raja~Ben Abdessalem, Shiva Nejati, Lionel~C Briand, and Thomas Stifter.
\newblock Testing vision-based control systems using learnable evolutionary algorithms.
\newblock In {\em 2018 IEEE/ACM 40th International Conference on Software Engineering (ICSE)}, pages 1016--1026. IEEE, 2018.

\bibitem{Arellano2024}
Silvia Arellano, Beatriz Otero, Tomasz~Piotr Kucner, and Ramon Canal.
\newblock A 3d terrain generator: Enhancing robotics simulations with gans.
\newblock In Giuseppe Nicosia, Varun Ojha, Emanuele La~Malfa, Gabriele La~Malfa, Panos~M. Pardalos, and Renato Umeton, editors, {\em Machine Learning, Optimization, and Data Science}, pages 212--226, Cham, 2024. Springer Nature Switzerland.

\bibitem{SAFE}
Mohammed Attaoui, Hazem Fahmy, Fabrizio Pastore, and Lionel Briand.
\newblock Black-box safety analysis and retraining of dnns based on feature extraction and clustering.
\newblock {\em ACM Trans. Softw. Eng. Methodol.}, 32(3), apr 2023.

\bibitem{attaoui2020multi}
Mohammed~Oualid Attaoui, Hanene Azzag, Mustapha Lebbah, and Nabil Keskes.
\newblock Multi-objective data stream clustering.
\newblock In {\em Proceedings of the 2020 Genetic and Evolutionary Computation Conference Companion}, pages 113--114, 2020.

\bibitem{Mohammed:PipelinesAssessment}
Mohammed~Oualid Attaoui, Hazem Fahmy, Fabrizio Pastore, and Lionel Briand.
\newblock Supporting safety analysis of image-processing dnns through clustering-based approaches.
\newblock {\em ACM Trans. Softw. Eng. Methodol.}, 33(5), jun 2024.

\bibitem{beamng}
{BeamNG GmbH}.
\newblock Beamng.drive.
\newblock \url{https://www.beamng.com}, 2015.

\bibitem{pymoo}
J.~{Blank} and K.~{Deb}.
\newblock pymoo: Multi-objective optimization in python.
\newblock {\em IEEE Access}, 8:89497--89509, 2020.

\bibitem{Osen}
Andreas Brands{\ae}ter and Ottar~L Osen.
\newblock Assessing autonomous ship navigation using bridge simulators enhanced by cycle-consistent adversarial networks.
\newblock {\em Proceedings of the Institution of Mechanical Engineers, Part O: Journal of Risk and Reliability}, 237(2):508--517, 2023.

\bibitem{carla}
{CARLA Simulator Team}.
\newblock Carla: An open urban driving simulator.
\newblock \url{https://github.com/carla-simulator/carla}, 2017.

\bibitem{chen2017deeplab}
Liang-Chieh Chen, George Papandreou, Iasonas Kokkinos, Kevin Murphy, and Alan~L. Yuille.
\newblock { DeepLab: Semantic Image Segmentation with Deep Convolutional Nets, Atrous Convolution, and Fully Connected CRFs}.
\newblock {\em IEEE Transactions on Pattern Analysis \& Machine Intelligence}, 40(04):834--848, April 2018.

\bibitem{chen2017rethinking}
Liang-Chieh Chen, George Papandreou, Florian Schroff, and Hartwig Adam.
\newblock Rethinking atrous convolution for semantic image segmentation.
\newblock {\em arXiv preprint arXiv:1706.05587}, 2017.

\bibitem{chen2018encoder}
Liang-Chieh Chen, Yukun Zhu, George Papandreou, Florian Schroff, and Hartwig Adam.
\newblock Encoder-decoder with atrous separable convolution for semantic image segmentation.
\newblock In {\em Proceedings of the European conference on computer vision (ECCV)}, pages 801--818, 2018.

\bibitem{Donkeycar}
Donkeycar community.
\newblock Donkeycar.
\newblock \url{https://www.donkeycar.com/}, 2024.

\bibitem{Dola2021}
Swaroopa Dola, Matthew~B. Dwyer, and Mary~Lou Soffa.
\newblock {Distribution-aware testing of neural networks using generative models}.
\newblock {\em Proceedings - International Conference on Software Engineering}, pages 226--237, 2021.

\bibitem{DONG2023103456}
Yuxuan Dong, Peng Wu, Sen Wang, and Yuanchang Liu.
\newblock Shipgan: Generative adversarial network based simulation-to-real image translation for ships.
\newblock {\em Applied Ocean Research}, 131:103456, 2023.

\bibitem{unrealengine}
{Epic Games}.
\newblock Unreal engine.

\bibitem{everingham2010pascal}
Mark Everingham, Luc Van~Gool, Christopher~KI Williams, John Winn, and Andrew Zisserman.
\newblock The pascal visual object classes (voc) challenge.
\newblock {\em International journal of computer vision}, 88(2):303--338, 2010.

\bibitem{fahmysupporting}
Hazem Fahmy, Fabrizio Pastore, Mojtaba Bagherzadeh, and Lionel Briand.
\newblock Supporting deep neural network safety analysis and retraining through heatmap-based unsupervised learning.
\newblock {\em IEEE Transactions on Reliability}, pages 1--17, 2021.

\bibitem{Hazem:SEDE}
Hazem Fahmy, Fabrizio Pastore, Lionel Briand, and Thomas Stifter.
\newblock Simulator-based explanation and debugging of hazard-triggering events in dnn-based safety-critical systems.
\newblock {\em ACM Trans. Softw. Eng. Methodol.}, 32(4), may 2023.

\bibitem{GambiASE19}
A.~{Gambi}, M.~{Mueller}, and G.~{Fraser}.
\newblock Asfault: Testing self-driving car software using search-based procedural content generation.
\newblock In {\em 2019 IEEE/ACM 41st International Conference on Software Engineering: Companion Proceedings (ICSE-Companion)}, pages 27--30, May 2019.

\bibitem{Gambi2019}
Alessio Gambi, Marc Mueller, and Gordon Fraser.
\newblock {Automatically testing self-driving cars with search-based procedural content generation}.
\newblock {\em ISSTA 2019 - Proceedings of the 28th ACM SIGSOFT International Symposium on Software Testing and Analysis}, pages 273--283, 2019.

\bibitem{Gladisch2019}
Christoph Gladisch, Thomas Heinz, Christian Heinzemann, Jens Oehlerking, Anne von Vietinghoff, and Tim Pfitzer.
\newblock Experience paper: search-based testing in automated driving control applications.
\newblock In {\em Proceedings of the 34th IEEE/ACM International Conference on Automated Software Engineering}, ASE '19, page 26–37. IEEE Press, 2020.

\bibitem{gluoncvnlp2020}
Jian Guo, He~He, Tong He, Leonard Lausen, Mu~Li, Haibin Lin, Xingjian Shi, Chenguang Wang, Junyuan Xie, Sheng Zha, Aston Zhang, Hang Zhang, Zhi Zhang, Zhongyue Zhang, Shuai Zheng, and Yi~Zhu.
\newblock Gluoncv and gluonnlp: Deep learning in computer vision and natural language processing.
\newblock {\em Journal of Machine Learning Research}, 21(23):1--7, 2020.

\bibitem{han2022fuzzgan}
Ge~Han, Zheng Li, Peng Tang, Chengyu Hu, and Shanqing Guo.
\newblock Fuzzgan: A generation-based fuzzing framework for testing deep neural networks.
\newblock In {\em 2022 IEEE 24th Int Conf on High Performance Computing \& Communications; 8th Int Conf on Data Science \& Systems; 20th Int Conf on Smart City; 8th Int Conf on Dependability in Sensor, Cloud \& Big Data Systems \& Application (HPCC/DSS/SmartCity/DependSys)}, pages 1601--1608. IEEE, 2022.

\bibitem{MORLOT}
F.~Ul Haq, D.~Shin, and L.~C. Briand.
\newblock Many-objective reinforcement learning for online testing of dnn-enabled systems.
\newblock In {\em 2023 IEEE/ACM 45th International Conference on Software Engineering (ICSE)}, pages 1814--1826, Los Alamitos, CA, USA, may 2023. IEEE Computer Society.

\bibitem{Haq:2021}
Fitash~Ul Haq, Donghwan Shin, Lionel~C. Briand, Thomas Stifter, and Jun Wang.
\newblock Automatic test suite generation for key-points detection dnns using many-objective search (experience paper).
\newblock In {\em Proceedings of the 30th ACM SIGSOFT International Symposium on Software Testing and Analysis}, ISSTA 2021, page 91–102, New York, NY, USA, 2021. Association for Computing Machinery.

\bibitem{fitash:offline:emse}
Fitash~Ul Haq, Donghwan Shin, Shiva Nejati, and Lionel Briand.
\newblock Can offline testing of deep neural networks replace their online testing? a case study of automated driving systems.
\newblock {\em Empirical Softw. Engg.}, 26(5), sep 2021.

\bibitem{Canada}
Fabrice Harel-Canada, Lingxiao Wang, Muhammad~Ali Gulzar, Quanquan Gu, and Miryung Kim.
\newblock Is neuron coverage a meaningful measure for testing deep neural networks?
\newblock In {\em Proceedings of the 28th ACM Joint Meeting on European Software Engineering Conference and Symposium on the Foundations of Software Engineering}, ESEC/FSE 2020, page 851–862, New York, NY, USA, 2020. Association for Computing Machinery.

\bibitem{he2016deep}
Kaiming He, Xiangyu Zhang, Shaoqing Ren, and Jian Sun.
\newblock Deep residual learning for image recognition.
\newblock In {\em Proceedings of the IEEE conference on computer vision and pattern recognition}, pages 770--778, 2016.

\bibitem{MUNIT}
Xun Huang, Ming-Yu Liu, Serge Belongie, and Jan Kautz.
\newblock Multimodal unsupervised image-to-image translation.
\newblock In {\em ECCV}, 2018.

\bibitem{isola2017image}
Phillip Isola, Jun-Yan Zhu, Tinghui Zhou, and Alexei~A Efros.
\newblock Image-to-image translation with conditional adversarial networks.
\newblock In {\em Proceedings of the IEEE conference on computer vision and pattern recognition}, pages 1125--1134, 2017.

\bibitem{SINVAD:Yoo:Feldt}
Sungmin Kang, Robert Feldt, and Shin Yoo.
\newblock Sinvad: Search-based image space navigation for dnn image classifier test input generation.
\newblock In {\em Proceedings of the IEEE/ACM 42nd International Conference on Software Engineering Workshops}, ICSEW'20, page 521–528, New York, NY, USA, 2020. Association for Computing Machinery.

\bibitem{li2021testing}
Zhong Li, Minxue Pan, Tian Zhang, and Xuandong Li.
\newblock Testing dnn-based autonomous driving systems under critical environmental conditions.
\newblock In {\em International Conference on Machine Learning}, pages 6471--6482. PMLR, 2021.

\bibitem{Liu}
Dingyu Liu, Yusheng Wang, Yonghoon Ji, Hiroshi Tsuchiya, Atsushi Yamashita, and Hajime Asama.
\newblock Cyclegan-based realistic image dataset generation for forward-looking sonar.
\newblock {\em Advanced Robotics}, 35(3-4):242--254, 2021.

\bibitem{Liu:UNIT}
Ming-Yu Liu, Thomas Breuel, and Jan Kautz.
\newblock Unsupervised image-to-image translation networks.
\newblock In {\em Proceedings of the 31st International Conference on Neural Information Processing Systems}, NIPS'17, page 700–708, Red Hook, NY, USA, 2017. Curran Associates Inc.

\bibitem{liu2024image}
Yanyan Liu, Xiaotian Bai, Jiafei Wang, Guoning Li, Jin Li, and Zengming Lv.
\newblock Image semantic segmentation approach based on deeplabv3 plus network with an attention mechanism.
\newblock {\em Engineering Applications of Artificial Intelligence}, 127:107260, 2024.

\bibitem{airsim1.8.1}
Microsoft.
\newblock Airsim.
\newblock \url{https://github.com/microsoft/AirSim}, 2024.
\newblock Version 1.8.1.

\bibitem{pix2pixHD}
NVIDIA.
\newblock pix2pixhd: High-resolution image synthesis with conditional gans.
\newblock \url{https://github.com/NVIDIA/pix2pixHD}, 2017.
\newblock 2023.

\bibitem{ACGAN}
Augustus Odena, Christopher Olah, and Jonathon Shlens.
\newblock Conditional image synthesis with auxiliary classifier gans.
\newblock In {\em Proceedings of the 34th International Conference on Machine Learning - Volume 70}, ICML'17, page 2642–2651. JMLR.org, 2017.

\bibitem{pmlr-v97-odena19a}
Augustus Odena, Catherine Olsson, David Andersen, and Ian Goodfellow.
\newblock {T}ensor{F}uzz: Debugging neural networks with coverage-guided fuzzing.
\newblock In Kamalika Chaudhuri and Ruslan Salakhutdinov, editors, {\em Proceedings of the 36th International Conference on Machine Learning}, volume~97 of {\em Proceedings of Machine Learning Research}, pages 4901--4911. PMLR, 09--15 Jun 2019.

\bibitem{pei2017deepxplore}
Kexin Pei, Yinzhi Cao, Junfeng Yang, and Suman Jana.
\newblock Deepxplore: Automated whitebox testing of deep learning systems.
\newblock In {\em proceedings of the 26th Symposium on Operating Systems Principles}, pages 1--18, 2017.

\bibitem{Riccio2021}
Vincenzo Riccio, Nargiz Humbatova, Gunel Jahangirova, and Paolo Tonella.
\newblock {DeepMetis: Augmenting a Deep Learning Test Set to Increase its Mutation Score}.
\newblock {\em Proceedings - 2021 36th IEEE/ACM International Conference on Automated Software Engineering, ASE 2021}, pages 355--367, 2021.

\bibitem{riccio2020model}
Vincenzo Riccio and Paolo Tonella.
\newblock {Model-based exploration of the frontier of behaviours for deep learning system testing}.
\newblock In {\em Proceedings of the 28th ACM Joint Meeting on European Software Engineering Conference and Symposium on the Foundations of Software Engineering}, number~2, pages 876--888, New York, NY, USA, nov 2020. ACM.

\bibitem{RiccioOOD}
Vincenzo Riccio and Paolo Tonella.
\newblock When and why test generators for deep learning produce invalid inputs: an empirical study.
\newblock In {\em 2023 IEEE/ACM 45th International Conference on Software Engineering (ICSE)}, pages 1161--1173, 2023.

\bibitem{sagar2021semantic}
Abhinav Sagar and RajKumar Soundrapandiyan.
\newblock Semantic segmentation with multi scale spatial attention for self driving cars.
\newblock In {\em Proceedings of the IEEE/CVF International Conference on Computer Vision}, pages 2650--2656, 2021.

\bibitem{shah2018airsim}
Shital Shah, Debadeepta Dey, Chris Lovett, and Ashish Kapoor.
\newblock Airsim: High-fidelity visual and physical simulation for autonomous vehicles.
\newblock In {\em Field and Service Robotics: Results of the 11th International Conference}, pages 621--635. Springer, 2018.

\bibitem{simonyan2015very}
Karen Simonyan and Andrew Zisserman.
\newblock Very deep convolutional networks for large-scale image recognition.
\newblock {\em arXiv preprint arXiv:1409.1556}, 2015.

\bibitem{PANGU}
{Star-Dundee}.
\newblock Planet and asteroid natural scene generation utility.
\newblock \url{https://pangu.software/}, 2024.
\newblock 2024.

\bibitem{stocco2022mind}
Andrea Stocco, Brian Pulfer, and Paolo Tonella.
\newblock Mind the gap! a study on the transferability of virtual vs physical-world testing of autonomous driving systems.
\newblock {\em IEEE Transactions on Software Engineering}, 2022.

\bibitem{swan2021ai4mars}
R~Michael Swan, Deegan Atha, Henry~A Leopold, Matthew Gildner, Stephanie Oij, Cindy Chiu, and Masahiro Ono.
\newblock Ai4mars: A dataset for terrain-aware autonomous driving on mars.
\newblock In {\em Proceedings of the IEEE/CVF conference on computer vision and pattern recognition}, pages 1982--1991, 2021.

\bibitem{Tian2017DeepTest}
Yuchi Tian, Kexin Pei, Suman Jana, and Baishakhi Ray.
\newblock Deeptest: Automated testing of deep-neural-network-driven autonomous cars.
\newblock In {\em 2018 IEEE/ACM 40th International Conference on Software Engineering (ICSE)}, pages 303--314. IEEE, 2017.

\bibitem{Ulbrich:Terminology}
Simon Ulbrich, Till Menzel, Andreas Reschka, Fabian Schuldt, and Markus Maurer.
\newblock Defining and substantiating the terms scene, situation, and scenario for automated driving.
\newblock In {\em 2015 IEEE 18th International Conference on Intelligent Transportation Systems}, pages 982--988, 2015.

\bibitem{VDA}
András Vargha and Harold~D. Delaney.
\newblock A critique and improvement of the cl common language effect size statistics of mcgraw and wong.
\newblock {\em Journal of Educational and Behavioral Statistics}, 25(2):101--132, 2000.

\bibitem{HIST}
Anton Vasiliuk, Daria Frolova, Mikhail Belyaev, and Boris Shirokikh.
\newblock Limitations of out-of-distribution detection in 3d medical image segmentation.
\newblock {\em Journal of Imaging}, 9(9), 2023.

\bibitem{wang2023dpnet}
Jun Wang, Xiaolin Zhang, Tianhong Yan, and Aihong Tan.
\newblock Dpnet: Dual-pyramid semantic segmentation network based on improved deeplabv3 plus.
\newblock {\em Electronics}, 12(14):3161, 2023.

\bibitem{wang2018pix2pixHD}
Ting-Chun Wang, Ming-Yu Liu, Jun-Yan Zhu, Andrew Tao, Jan Kautz, and Bryan Catanzaro.
\newblock High-resolution image synthesis and semantic manipulation with conditional gans.
\newblock In {\em 2018 IEEE/CVF Conference on Computer Vision and Pattern Recognition (CVPR)}, pages 8798--8807, Los Alamitos, CA, USA, jun 2018. IEEE Computer Society.

\bibitem{Wei:AdversarialStyle}
Jiefei Wei and Qinggang Meng.
\newblock Adversarialstyle: Gan based style guided verification framework for deep learning systems.
\newblock In {\em 2020 IEEE 18th International Conference on Industrial Informatics (INDIN)}, volume~1, pages 641--648, 2020.

\bibitem{AdversarialStyle}
Jiefei Wei and Qinggang Meng.
\newblock Adversarialstyle: Gan based style guided verification framework for deep learning systems.
\newblock In {\em 2020 IEEE 18th International Conference on Industrial Informatics (INDIN)}, volume~1, pages 641--648, 2020.

\bibitem{TonellaAmbiguity}
Michael Weiss, Andr\'{e}~Garc\'{\i}a G\'{o}mez, and Paolo Tonella.
\newblock Generating and detecting true ambiguity: a forgotten danger in dnn supervision testing.
\newblock {\em Empirical Softw. Engg.}, 28(6), November 2023.

\bibitem{woodlief2022semantic}
Trey Woodlief, Sebastian Elbaum, and Kevin Sullivan.
\newblock Semantic image fuzzing of ai perception systems.
\newblock In {\em Proceedings of the 44th International Conference on Software Engineering}, pages 1958--1969, 2022.

\bibitem{xie2019deephunter}
Xiaofei Xie, Lei Ma, Felix Juefei-Xu, Minhui Xue, Hongxu Chen, Yang Liu, Jianjun Zhao, Bo~Li, Jianxiong Yin, and Simon See.
\newblock Deephunter: a coverage-guided fuzz testing framework for deep neural networks.
\newblock In {\em Proceedings of the 28th ACM SIGSOFT International Symposium on Software Testing and Analysis}, pages 146--157, 2019.

\bibitem{SurfelGAN}
Z.~Yang, Y.~Chai, D.~Anguelov, Y.~Zhou, P.~Sun, D.~Erhan, S.~Rafferty, and H.~Kretzschmar.
\newblock Surfelgan: Synthesizing realistic sensor data for autonomous driving.
\newblock In {\em 2020 IEEE/CVF Conference on Computer Vision and Pattern Recognition (CVPR)}, pages 11115--11124, Los Alamitos, CA, USA, jun 2020. IEEE Computer Society.

\bibitem{ManifoldExploration}
Yuanyuan Yuan, Qi~Pang, and Shuai Wang.
\newblock Provably valid and diverse mutations of real-world media data for dnn testing.
\newblock {\em IEEE Transactions on Software Engineering}, 50(5):1040--1064, 2024.

\bibitem{zhang2018deeproad}
Mengshi Zhang, Yuqun Zhang, Lingming Zhang, Cong Liu, and Sarfraz Khurshid.
\newblock Deeproad: Gan-based metamorphic testing and input validation framework for autonomous driving systems.
\newblock In {\em Proceedings of the 33rd ACM/IEEE International Conference on Automated Software Engineering}, pages 132--142, 2018.

\bibitem{zhang2021cagfuzz}
Pengcheng Zhang, Bin Ren, Hai Dong, and Qiyin Dai.
\newblock Cagfuzz: coverage-guided adversarial generative fuzzing testing for image-based deep learning systems.
\newblock {\em IEEE Transactions on Software Engineering}, 48(11):4630--4646, 2021.

\bibitem{Zhang2023}
Xinhai Zhang, Jianbo Tao, Kaige Tan, Martin T{\"{o}}rngren, Jos{\'{e}} Manuel~Gaspar S{\'{a}}nchez, Muhammad~Rusyadi Ramli, Xin Tao, Magnus Gyllenhammar, Franz Wotawa, Naveen Mohan, Mihai Nica, and Hermann Felbinger.
\newblock {Finding Critical Scenarios for Automated Driving Systems: A Systematic Mapping Study}.
\newblock {\em IEEE Transactions on Software Engineering}, 49(3):991--1026, mar 2023.

\bibitem{Zhu:2017}
Jun-Yan Zhu, Taesung Park, Phillip Isola, and Alexei~A. Efros.
\newblock Unpaired image-to-image translation using cycle-consistent adversarial networks.
\newblock In {\em 2017 IEEE International Conference on Computer Vision (ICCV)}, pages 2242--2251, 2017.

\bibitem{zohdinasab2021deephyperion}
Tahereh Zohdinasab, Vincenzo Riccio, Alessio Gambi, and Paolo Tonella.
\newblock Deephyperion: exploring the feature space of deep learning-based systems through illumination search.
\newblock In {\em Proceedings of the 30th ACM SIGSOFT International Symposium on Software Testing and Analysis}, pages 79--90, 2021.

\bibitem{Zohdinasab2023}
Tahereh Zohdinasab, Vincenzo Riccio, Alessio Gambi, and Paolo Tonella.
\newblock Efficient and effective feature space exploration for testing deep learning systems.
\newblock {\em ACM Trans. Softw. Eng. Methodol.}, 32(2), mar 2023.

\bibitem{Zohdinasab2023b}
Tahereh Zohdinasab, Vincenzo Riccio, and Paolo Tonella.
\newblock {An Empirical Study on Low- and High-Level Explanations of Deep Learning Misbehaviours}.
\newblock {\em International Symposium on Empirical Software Engineering and Measurement}, pages 1--11, 2023.

\end{thebibliography}
